\documentclass[aip,pop,reprint]{revtex4-1} 

\usepackage{graphicx}
\usepackage{amssymb,amsmath}
\usepackage{bm}
\usepackage{siunitx}
\allowdisplaybreaks

\begin{document}

\title{Effects of electron temperature anisotropy on proton mirror instability evolution}
\author{Narges Ahmadi}
\author{Kai Germaschewski}
\author{Joachim Raeder}
\affiliation{University of New Hampshire, Durham, NH 03824}

\begin{abstract}
Proton mirror modes are large amplitude nonpropagating structures frequently
observed in the magnetosheath. It has been suggested that electron
temperature anisotropy can enhance the proton mirror instability growth rate
while leaving the proton cyclotron instability largely unaffected, therefore
causing the proton mirror instability to dominate the proton cyclotron
instability in Earth's magnetosheath. Here, we use particle-in-cell
simulations to investigate the electron temperature anisotropy
effects on proton  mirror instability evolution. Contrary to the hypothesis, electron temperature
anisotropy leads to excitement of the electron whistler instability. Our
results show that the electron whistler instability grows much faster
than the proton mirror instability and quickly consumes the electron free
energy, so that there is no electron temperature anisotropy left to
significantly impact the evolution of the proton mirror instability.
\end{abstract}
\maketitle

\section{Introduction}

There is a region of the dayside magnetosheath which is characterized
by temperature anisotropy $T_{\perp}/T_{\parallel}>1$, where $T_\perp$
and $T_\parallel$ indicates the perpendicular and parallel
temperatures relative to the background magnetic field $B_0$,
respectively. The temperature anisotropy is caused by plasma heating at the
quasi-perpendicular bow shock and field-line draping close to the
magnetopause as shown by Midgley and Davis \cite{midgleycalculation1963} and
by Zwan and Wolf \cite{zwandepletion1976}.
This temperature anisotropy leads to the generation of low frequency waves. For
$T_{p\perp}/T_{p\parallel} > 1$, where $T_p$ shows proton temperature,
proton cyclotron waves \cite{kennellimit1966} and proton
mirror waves \cite{chandrasekharthe1958, hasegawadrift1969} are generated, and for $T_{e\perp}/T_{e\parallel} > 1$,
where $T_e$ stands for electron temperature, electron whistler waves
\cite{kennellimit1966, scharercyclotron1967} and electron
mirror waves \cite{garylinear2006} can grow.
The proton cyclotron instability is a resonant instability and it
propagates parallel to the background magnetic field with frequencies
less than the proton gyrofrequency ($\omega < \Omega_p$) while the proton mirror
instability has zero frequency ($\omega = 0$) in the plasma frame and
its wave vector is
oblique to the background magnetic field. Here, $\Omega_p$ denotes
the proton gyrofrequency. The mirror instability creates
magnetic depressions or magnetic mirrors in the plasma which can trap
particles and in this way, particles exchange their kinetic energy to
the wave and instability grows, as shown by Southwood and Kivelson\cite{southwoodmirror1993}.
Linear theory predicts that mirror modes are more likely to be dominant in high $\beta_p$ regions
of the magnetosheath while proton cyclotron modes are dominant in the
low $\beta_p$ plasma conditions, where $\beta_p$ is the ratio between
parallel proton pressure to magnetic pressure.\cite{garythe1992}

The electron whistler waves
propagate parallel to the background magnetic field with frequencies
larger than proton gyrofrequency and smaller than electron
gyrofrequency ($\Omega_p < \omega < \Omega_e$), where $\Omega_e$  is the
electron gyrofrequency. The electron mirror
instability is similar to the proton mirror instability but its
wavelength is of the order of electron inertial length ($d_e =
c/\omega_{pe}$). All of these
instabilities compete with each other to consume the available free
energy of the system which is constrained in the temperature anisotropies.
Proton cyclotron and proton mirror instabilities compete with each
other for the available free energy in proton temperature anisotropy
while electron whistler and electron mirror instabilities compete for
consuming the electron temperature anisotropy. But there is also a
competition between proton mirror instability and electron whistler
instability to consume the available electron free energy which we are
interested to study in this paper.

There are frequent observations of proton mirror mode structures in
the Earth's magnetosheath.\cite{kaufmannlarge1970,tsurutanilion1982}
Proton mirror modes have also been observed in the solar
wind \cite{winterhalterulysses1994}, at comets\cite{russellmirror1987},
in the magnetosheaths of other planets like Jupiter and Saturn
\cite{erdos1996,cattaneoevolution1998}, and in the
heliosheath.\cite{burlagatrains2006} Proton mirror modes have been observed in
regions with low proton plasma beta $\beta_p$, although the linear dispersion theory
predicts that proton cyclotron mode should be the dominant mode in
these regions.
Price et al. \cite{pricenumerical1986} showed that the presence of heavy ions tends to
suppress the proton cyclotron instability while the growth rate of the
proton mirror instability is not significantly affected. This can be one
possible mechanism for proton mirror modes to dominate proton cyclotron
instability.\cite{garythe1992, garyion1993}

Shoji et al. \cite{shojimirror2009} performed two and three dimensional hybrid simulations to
study the competition between these two modes. They suggested that in
three dimensional simulations, proton mirror modes consume most of the free
energy of the system and it stops the growth of the proton cyclotron
waves. Porazik and Johnson \cite{poraziklinear2013} used the gyrokinetic theory to
derive the linear dispersion relation for the proton mirror
instability and provided a coherent view of different kinetic
approaches that is used to obtain the dispersion relation.

The purpose of this work is to study the effects of electron
temperature anisotropy on the evolution of proton mirror
instability. Linear dispersion theory shows that the
electron temperature anisotropy enhances the proton mirror instability growth
rate but it doesn't affect the proton cyclotron instability growth
rate significantly.\cite{garythe1992} Since we need to consider electron
dynamics, we use particle in cell simulations to include kinetic
effects of both protons and electrons.
Electrons get anisotropically heated in the shock layer similar to
protons \cite{burgess2012}. Some previous studies have assumed electrons to be isotropic, since
they performed hybrid simulations which treats electrons as a fluid
\cite{shojimirror2009,hellingermagnetosheath2005}. However, Tsurutani et al. \cite{tsurutanilion1982} have
shown that the electron temperature anisotropy is generally larger
than $1$ in Earth's magnetosheath. 

Masood et al. \cite{masoodobservations2008} analyzed Cluster data in Earth's magnetosheath
and found that electrons exhibit significant temperature
anisotropy in the deep magnetosheath due to magnetic field line
draping while being isotropic downstream
of the quasi-perpendicular bow shock.
Remya et al. \cite{remyaion2013} used linear theory to study the role 
of electron temperature anisotropy on the proton cyclotron and proton
mirror instabilities and they conclude that an inclusion of
anisotropic electrons with $T_{e\perp}/T_{e||} \ge 1.2$ reduces the
proton cyclotron growth rate substantially and increases the proton
mirror instability growth rate. But we need to mention that they are ignoring the electron whistler
instability presence.

In section 2, we solve the linear dispersion relation to find the growth
rates of the instabilities for given plasma parameters. In section 3,
we benchmark our kinetic code with linear dispersion theory for both
proton temperature anisotropy and electron temperature anisotropy instabilities.
In section 4, we present simulation results for different proton to electron mass
ratios and how electron anisotropy affects the growth of the proton
mirror instability. In section 5, we discuss the conclusions.

\section{Linear Analysis}

We solved the linear dispersion relation for a homogeneous,
collisionless plasma with bi-Maxwellian distributions
to measure the growth rates of the temperature anisotropy instabilities for typical
magnetosheath plasma parameters. We consider two species: protons and
electrons. We assume charge neutrality $n_p=n_e$ and zero
relative drift between the electrons and protons.\cite{stix1962}
Solutions of the linear dispersion equation are typically
expressed in terms of dimensionless variables. It is natural to use
electron inertial length and electron gyrofrequency as
normalizing factors for electrons and proton inertial length and
proton gyrofrequency for normalizing proton related
instabilities.

In Earth's magnetosheath, the distributions become anisotropic because of heating of the
particles across the quasi-perpendicular bow shock and field line draping. The time scale of
the heating through the shock is about one proton gyroperiod.
This time scale is very fast and does not allow the proton
instabilities to grow in the shock layer. Therefore a
considerable amount of proton temperature anisotropy is left downstream of the
quasi-perpendicular shock in
the magnetosheath. For electrons, on the other hand, one proton gyroperiod
equals 1836 electron gyroperiods. Thus,
electron instabilities have sufficient time to grow and isotropize the
electron distributions. Therefore, we consider high
proton temperature anisotropies and lower electron temperature
anisotropies to resemble the magnetosheath plasma conditions
downstream of the quasi-perpendicular shock.\cite{burgess2012}

\subsection{Competition between electron whistler and electron mirror instability}
\begin{figure}
  \centering
  \includegraphics[width=\columnwidth]{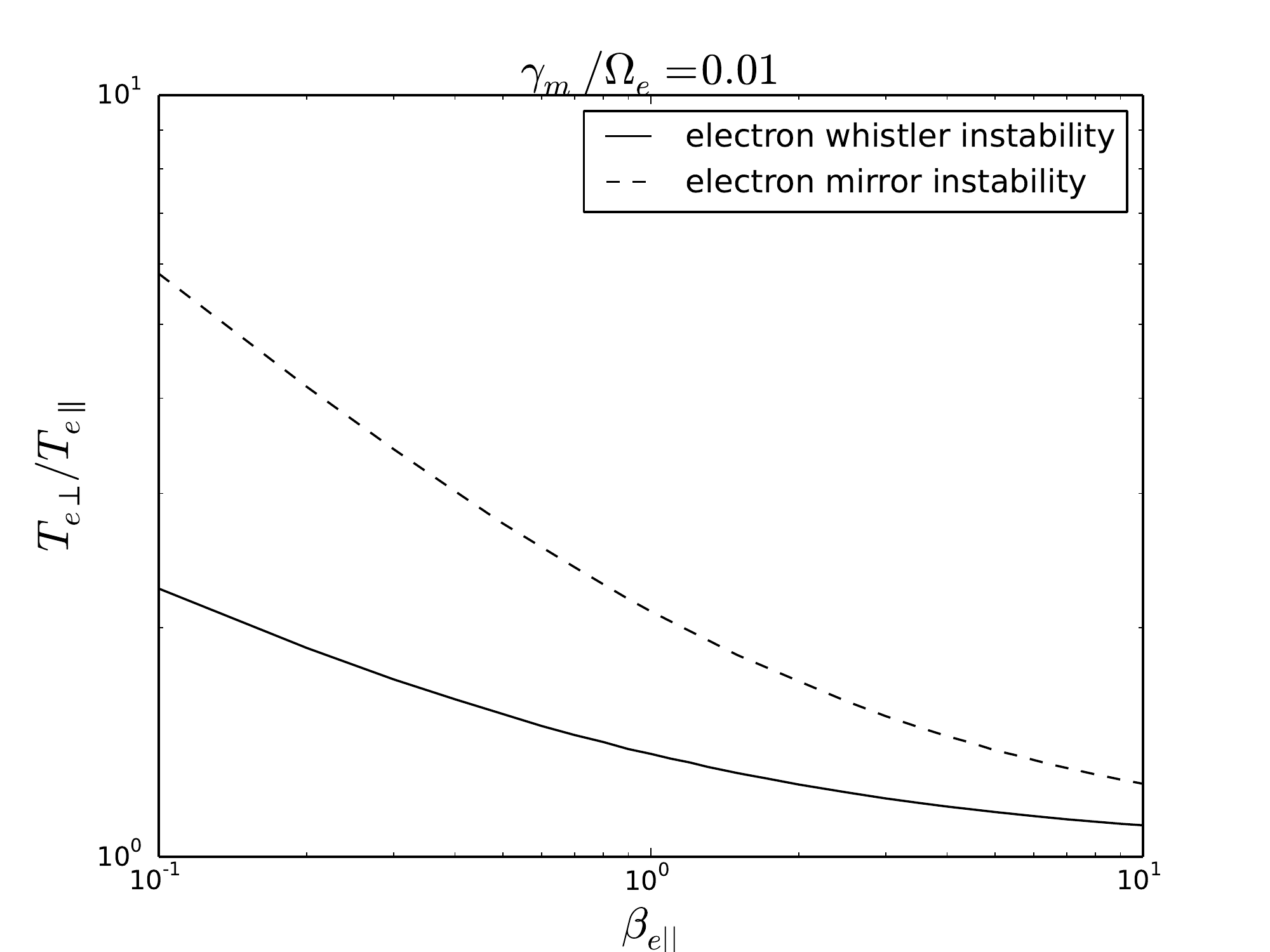}
  \caption{Electron temperature anisotropy at the $\gamma_m/\Omega_e =
    0.01$ thresholds of electron whistler and electron mirror
    instabilities as function of $\beta_{e||}$. The solid line
    shows the instability threshold of electron whistler
    instability and the dashed line shows the electron mirror
    instability threshold. If the plasma parameters lie below the
    threshold line, the instabilities won't be able to grow.}
  \label{fig:electron_thresholds}
\end{figure}

Figure \ref{fig:electron_thresholds} shows the instability thresholds
for electron whistler and electron mirror instabilities. We keep
$T_{p\perp}/T_{p||} = 1$ and $\beta_p=1$. The instability thresholds
($\gamma_m/\Omega_e = 0.01$)
are measured using linear dispersion theory. $\gamma_m$ refers to
maximum growth rate. Comparing
the electron whistler and electron mirror instability growth rates in
Figure \ref{fig:electron_thresholds}, we clearly see
that the electron whistler instability has a lower instability
threshold than the electron
mirror instability and it may therefore suppress the electron mirror
mode. Observations show that electrons follow the marginal stability
path of the electron whistler instability in Earth's magnetosheath
which indicates that
electron whistler instability is the dominant instability.\cite{garyelectron2005}

\subsection{Competition between proton cyclotron and proton mirror instability}
\begin{figure}
  \centering
  \includegraphics[width=\columnwidth]{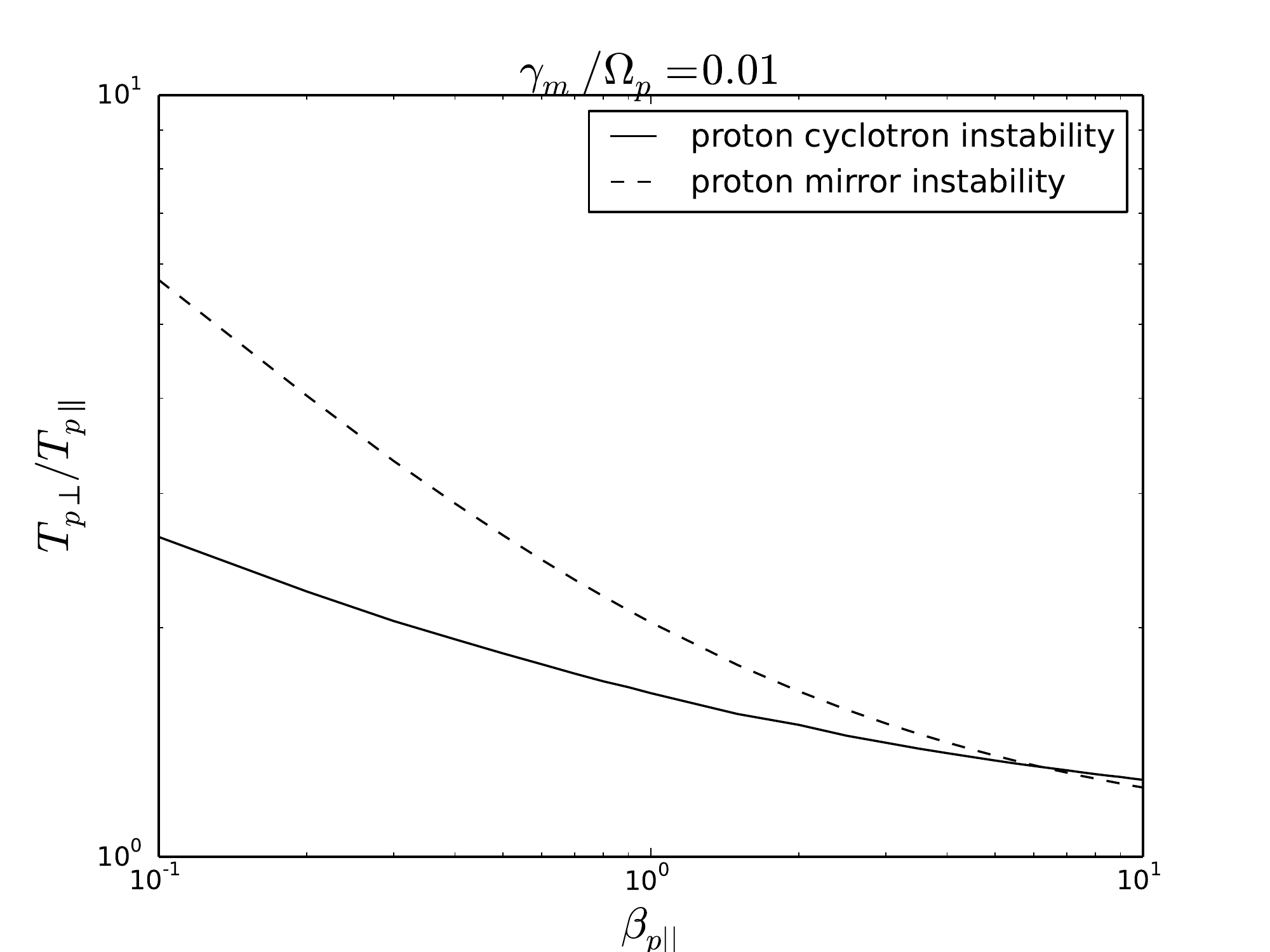}
  \caption{Proton temperature anisotropy at the $\gamma_m/\Omega_p =
    0.01$ thresholds of proton cyclotron and proton mirror
    instabilities as function of $\beta_{p||}$. The solid line
    shows the instability threshold for proton cyclotron
    instability and the dashed line shows the proton mirror
    instability threshold. If the plasma parameters lie below the
    threshold line, the instabilities won't be able to grow.}
  \label{fig:proton_thresholds}
\end{figure}

In the case of the proton temperature anisotropy instabilities, the proton cyclotron instability has
larger growth rate compared to the proton mirror instability for low
proton plasma beta $\beta_p$ and it should be the dominant instability
in the magnetosheath as shown in Figure
\ref{fig:proton_thresholds}. In Figure \ref{fig:proton_thresholds}, we
keep electrons isotropic and measure the proton cyclotron and mirror
instability thresholds ($\gamma_m/\Omega_p = 0.01$) using linear
dispersion theory. It is clear that the proton cyclotron instability has larger
growth rate compared to mirror instability for low
$\beta_p$ and high $T_{p\perp}/T_{p\parallel}$.
But observations show that in regions where
we expect the dominance of the proton cyclotron instability, mirror
instability has grown and it is the dominant mode. So the question is
what helps the proton mirror instability to grow faster than the proton cyclotron
instability in low $\beta_p$ regions?

\begin{figure}
  \centering
  \includegraphics[width=\columnwidth]{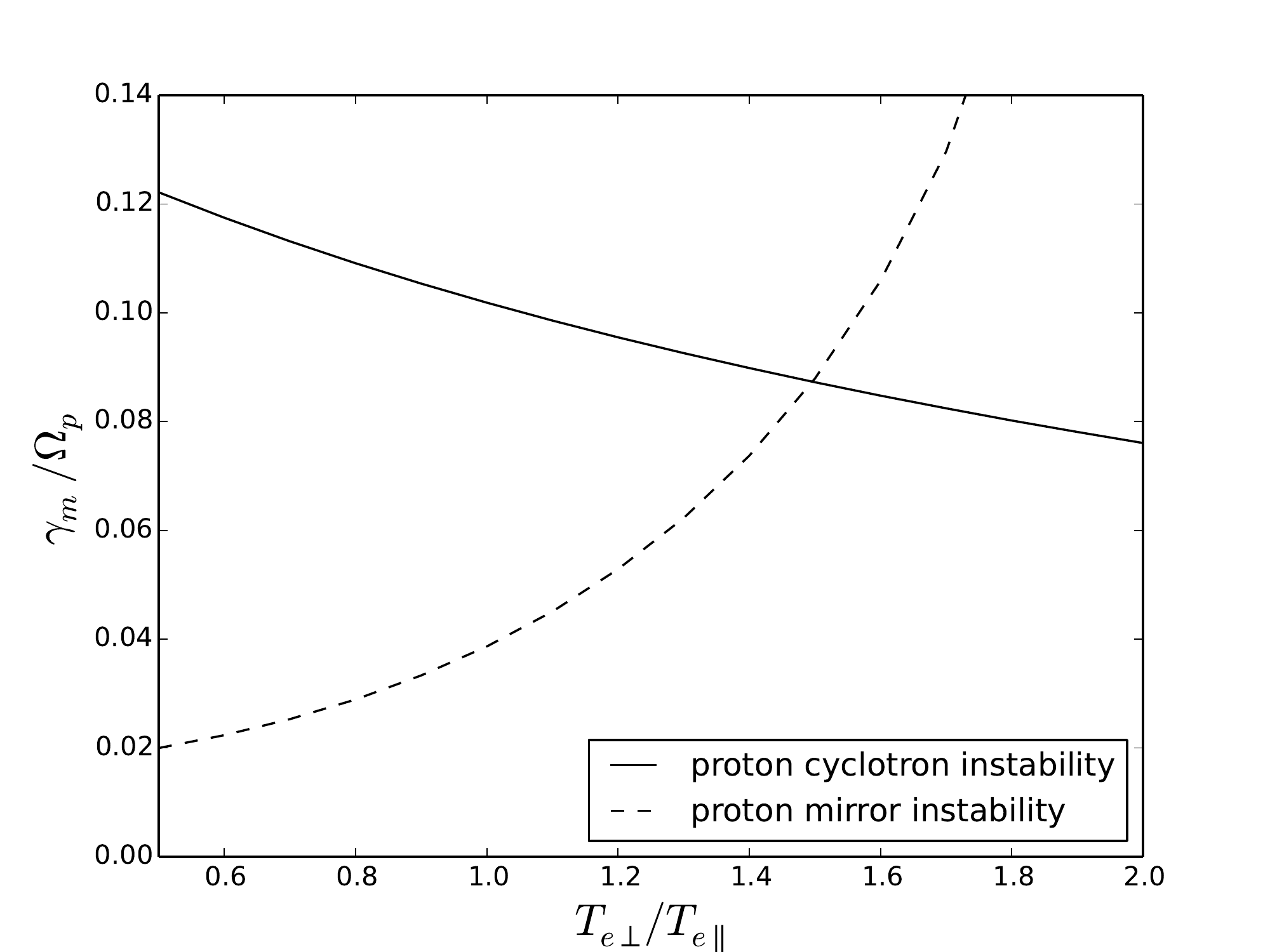}
  \caption{Electron temperature anisotropy effects on mirror
   instability and proton cyclotron maximum growth rates. Solid line
   shows the maximum growth rate of the proton cyclotron instability
   as a function of electron temperature anisotropy and dashed line
   shows the maximum growth rate of proton mirror
   instability. $T_{p\perp}/T_{p\parallel} = 2.5$ and
   $\beta_p=\beta_e=1$ are fixed. }
 \label{fig:electron_anisotropy}
\end{figure}

One possibility is the
effects of electron temperature anisotropy on the proton mirror instability growth
rate. Figure \ref{fig:electron_anisotropy} shows that by increasing the electron temperature anisotropy,
mirror instability growth rate increases while leaving the proton
cyclotron instability only slightly affected. The reason is that proton cyclotron
instability is a resonant instability and electrons do not
resonate with the proton cyclotron mode, but they can get trapped in the
magnetic bottles of the mirror instability and exchange energy with the
wave. 

In order to study the nonlinear effects of the electron dynamics on the evolution of
the proton mirror instability, we use particle in cell simulations. 

\begin{figure}
  \centering
  \includegraphics[width=\columnwidth]{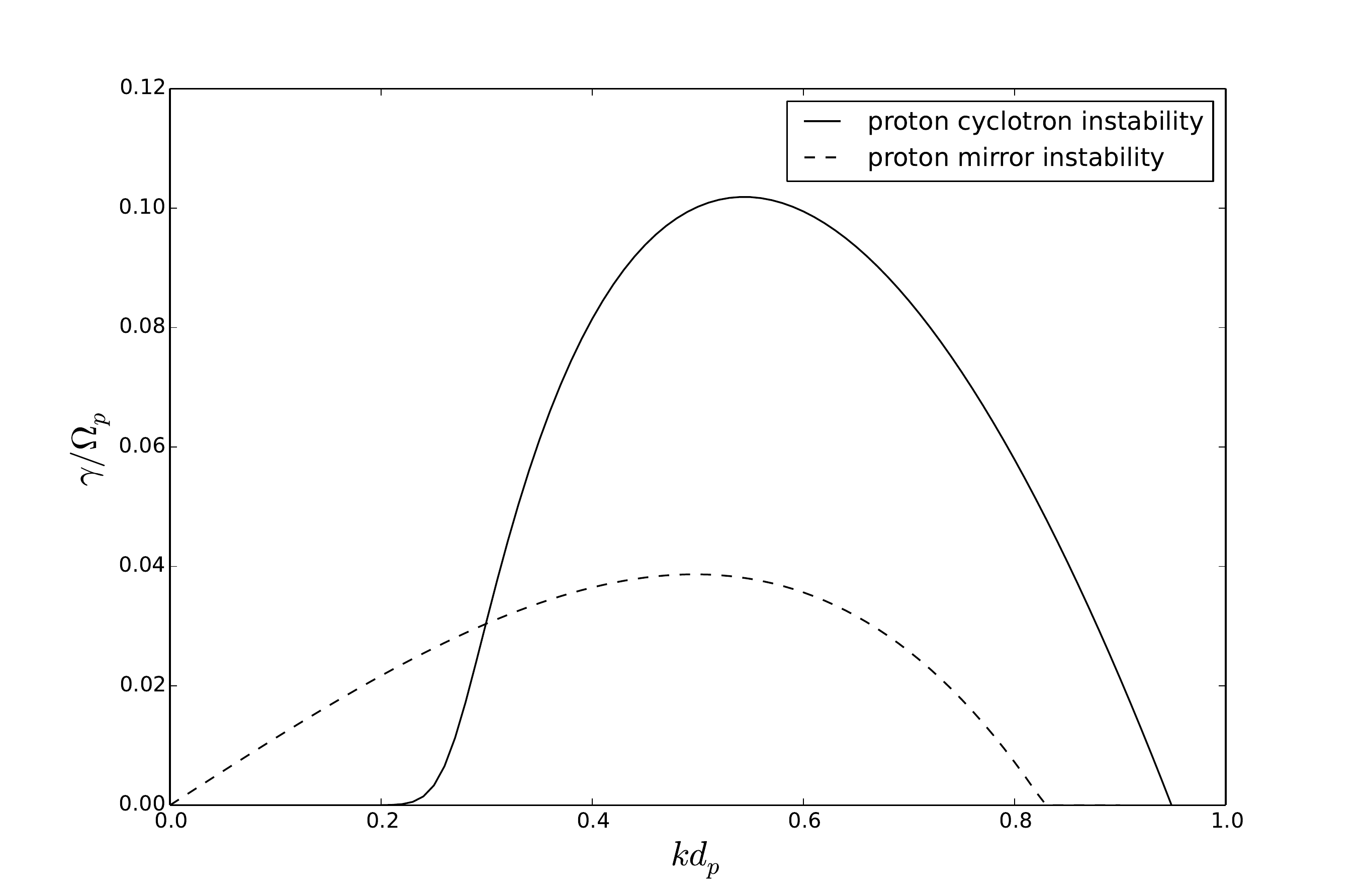}
  \caption{The growth rate as a function of wave number for
    proton cyclotron and proton mirror instability. Solid line shows
    the proton cyclotron instability growth rate at $\theta_m=0$ while
    the dashed line shows the growth rate of proton mirror
    instability at $\theta_m=63$. The maximum growth rate of proton
    cyclotron instability is $\gamma_m/\Omega_p = 0.1$ at $k_m d_p =
    0.54$ while the proton mirror instability maximum growth rate is
    $\gamma_m/\Omega_p = 0.039$ with $k_m d_p = 0.5$.}
  \label{fig:growthrates}
\end{figure} 

\section{Verification of \texttt{PSC} by comparison to linear dispersion theory}

\texttt{PSC} is a state of the art electromagnetic particle-in-cell
simulations code described by Germascheski et al.\cite{germaschewski2013} In this
section, we compare the \texttt{PSC} results with linear dispersion theory. In
order to show that \texttt{PSC} can capture temperature anisotropy
instabilities correctly, we measured the growth rate of the instabilities from
simulation in the linear regime for selected plasma parameters and compare with linear
theory predictions. We start with bi-Maxwellian protons and Maxwellian
electrons. We choose $T_{p\perp}/T_{p\parallel} =2.5$,
$T_{e\perp}/T_{e\parallel} =1$, $\beta_p = \beta_e = 1$. We perform
two-dimensional simulations with $L_y=L_z=128d_p$ where $L_y$ and
$L_z$ are the length of the simulation box in $y$ and $z$
directions, $\omega_p$ is the proton plasma frequency and $d_p=c/\omega_p$
is the proton inertial length. The number of grid points $(n_y\times n_z)$ is
$4096\times 4096$. Periodic boundary conditions are used in both dimensions. A
constant background magnetic field $\mathbf{B_0}$ is assumed in the $z$ direction. 

\begin{figure}
  \centering
  \includegraphics[width=\columnwidth]{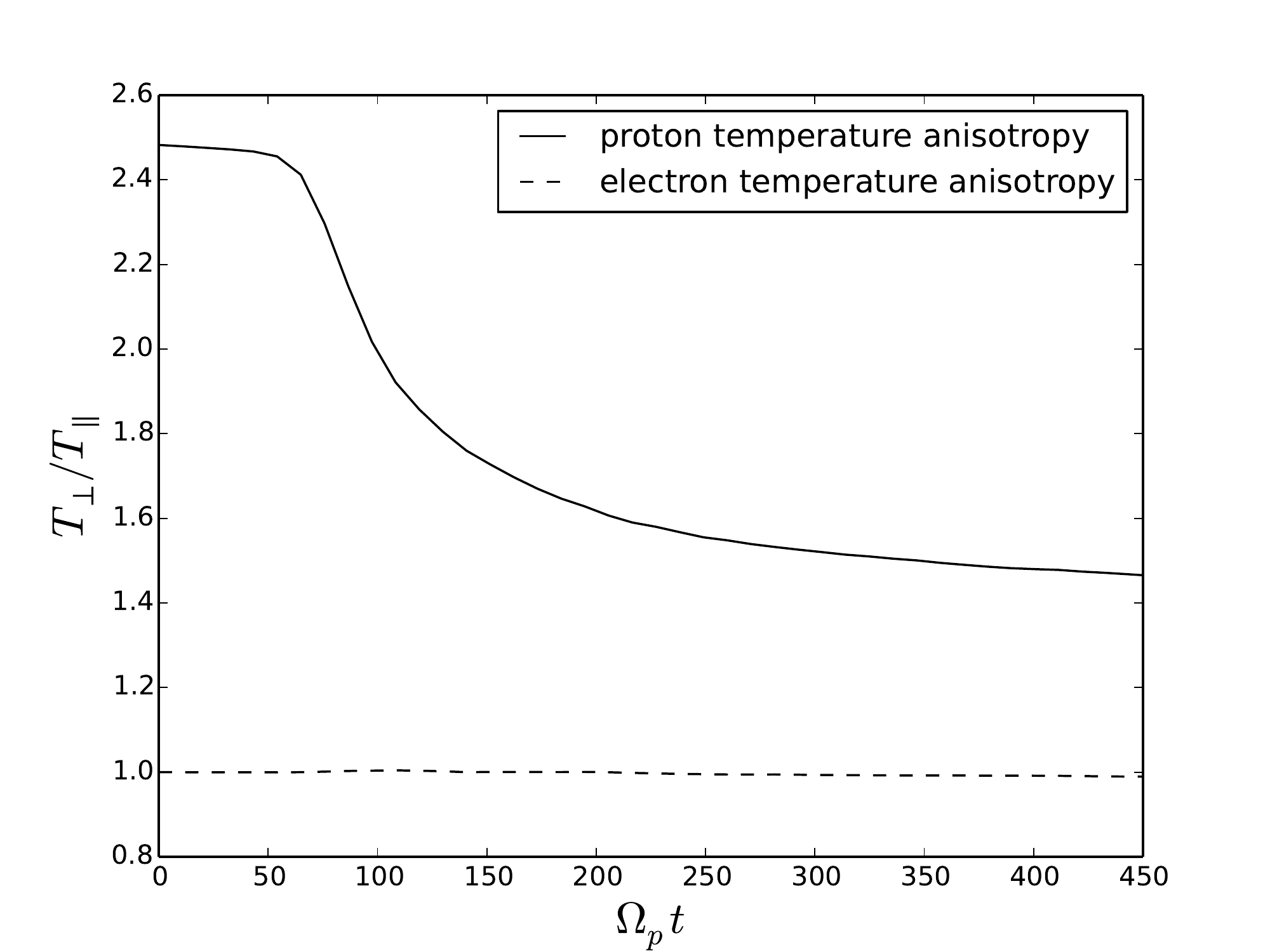}
  \caption{Proton and electron temperature anisotropy evolution
    as a function of time in 2D particle in cell simulation. Initial
    parameters are $T_{p\perp}/T_{p\parallel} = 2.5$,
    $T_{e\perp}/T_{e\parallel} = 1$ and $\beta_p=\beta_e=1$. The linear
    regime of proton temperature anisotropy instabilities is about
    $\Omega_p t = 70$.}
  \label{fig:aniso_benchmark}
\end{figure}

\begin{figure}
  \centering
  \includegraphics[width=\columnwidth]{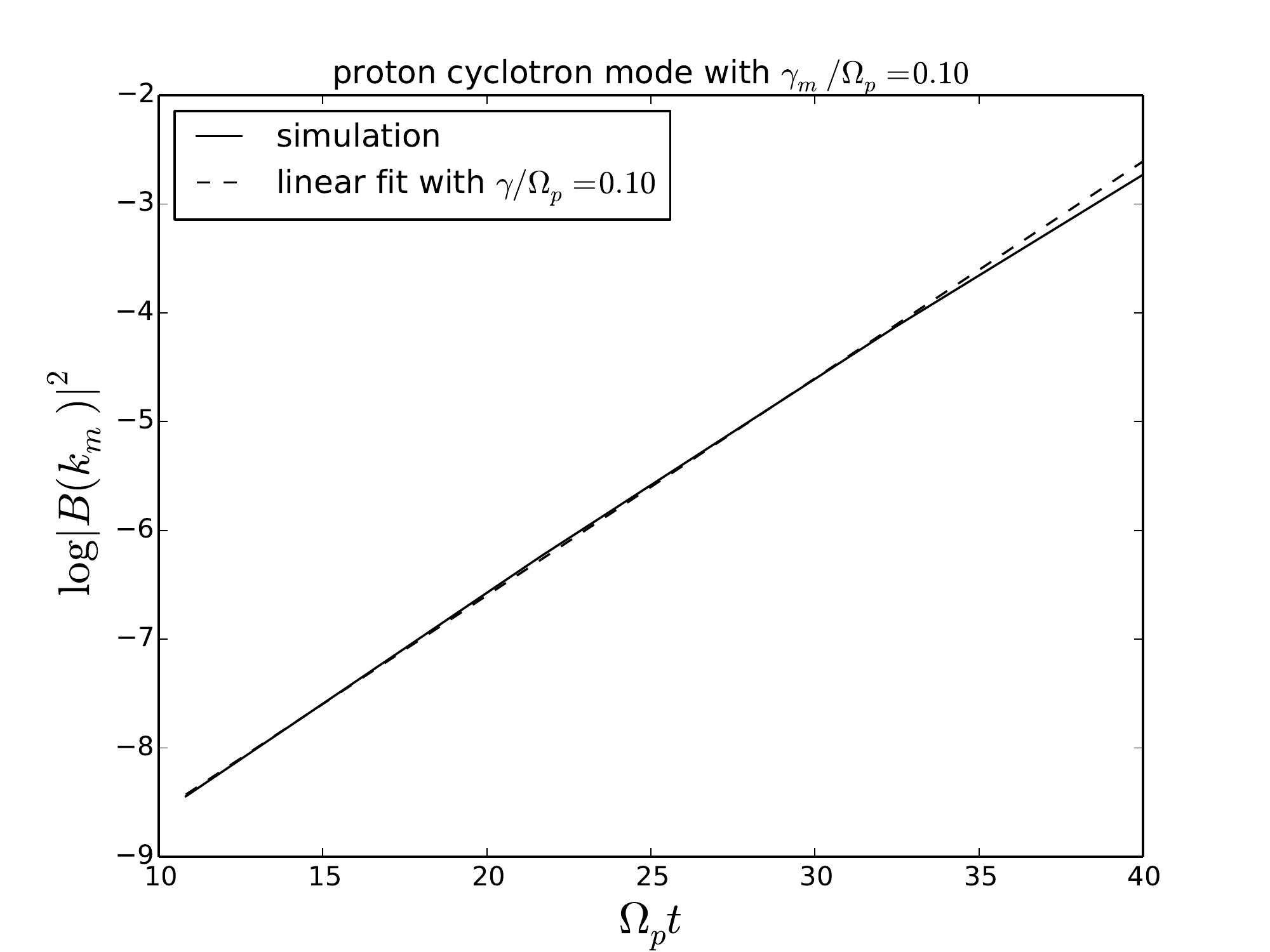}
  \caption{Measured maximum growth rate of proton cyclotron instability from
    simulation in the linear regime. The measured growth rate is in agreement with linear
    dispersion theory prediction. }
  \label{fig:protonfit}
\end{figure}

\begin{figure}
  \centering
  \includegraphics[width=\columnwidth]{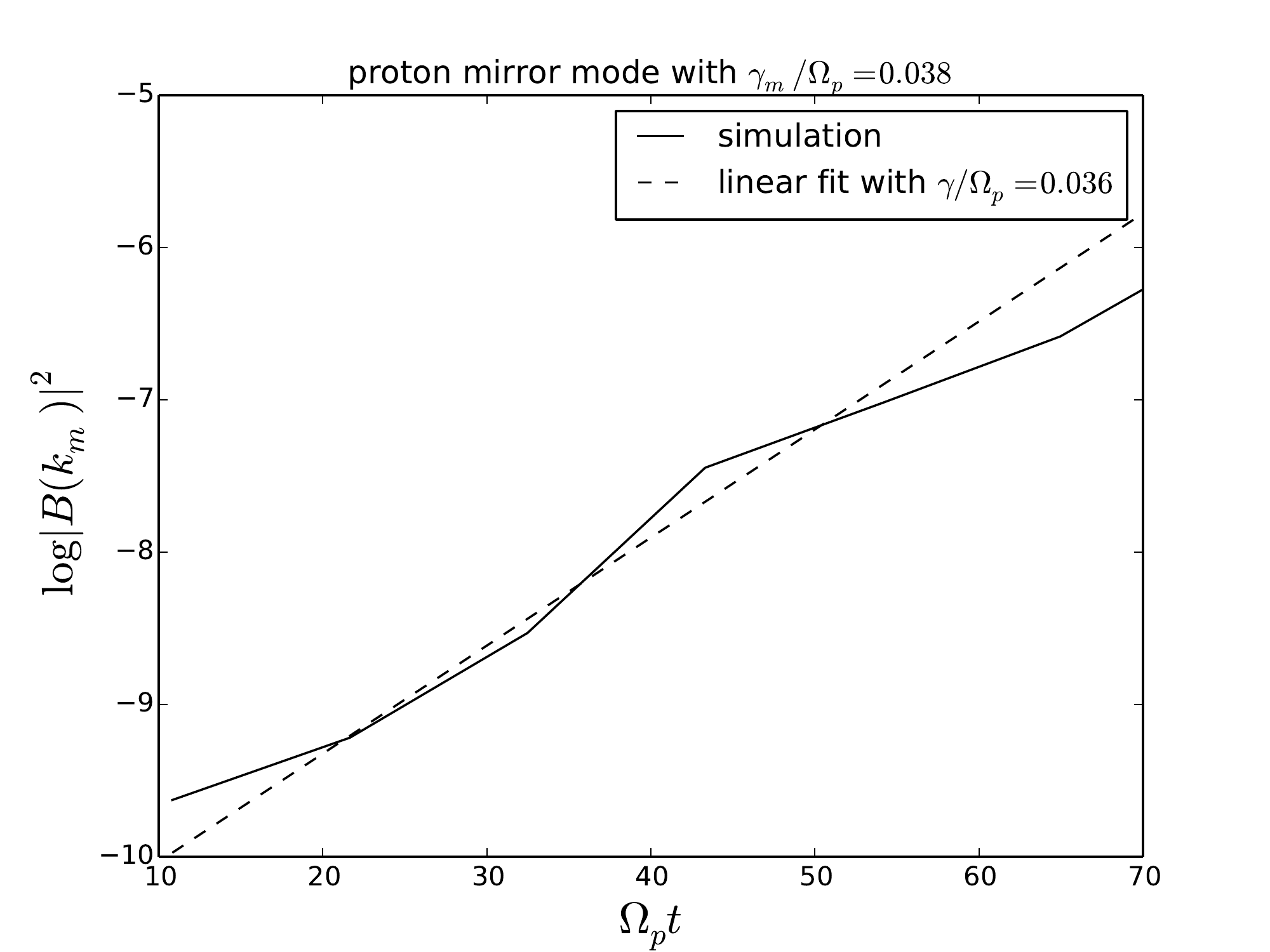}
  \caption{Measured maximum growth rate of proton mirror instability from
    simulation in the linear regime. The measured growth rate is in good agreement with linear
    dispersion theory prediction.}
  \label{fig:mirrorfit}
\end{figure}

 With anisotropic protons $(T_{p\perp}/T_{p||}>1)$, proton cyclotron and proton mirror instabilities
will grow. From linear theory, we expect the maximum growth rate of
the proton
cyclotron instability to be $\gamma_m= 0.10\Omega_p$ at $k_m d_p =
0.54$ and $\theta=0^\circ$ while the proton mirror instability maximum growth rate is
 $\gamma_m= 0.039\Omega_p$ with $k_m d_p = 0.50$ at 
$\theta =63^\circ$ as shown in Figure \ref{fig:growthrates}. $\theta$ is the angle
between the wave number vector $\mathbf{k}$ and $\mathbf{B_0}$. Figure
\ref{fig:aniso_benchmark} shows the temperature anisotropy evolution
of both protons and electrons. Electrons remain isotropic. As proton cyclotron and proton mirror
instabilities start growing, the proton temperature anisotropy
decreases. The linear regime of the proton temperature anisotropy
instabilities extends through about $\Omega_p t= 70$ in this case. Figures
\ref{fig:protonfit} and \ref{fig:mirrorfit} compare the measured
maximum growth rate of proton cyclotron and proton mirror instabilities from
simulation with linear dispersion theory predictions.
The simulation results are in good agreement with linear theory.

\begin{figure}
  \centering
  \includegraphics[width=\columnwidth]{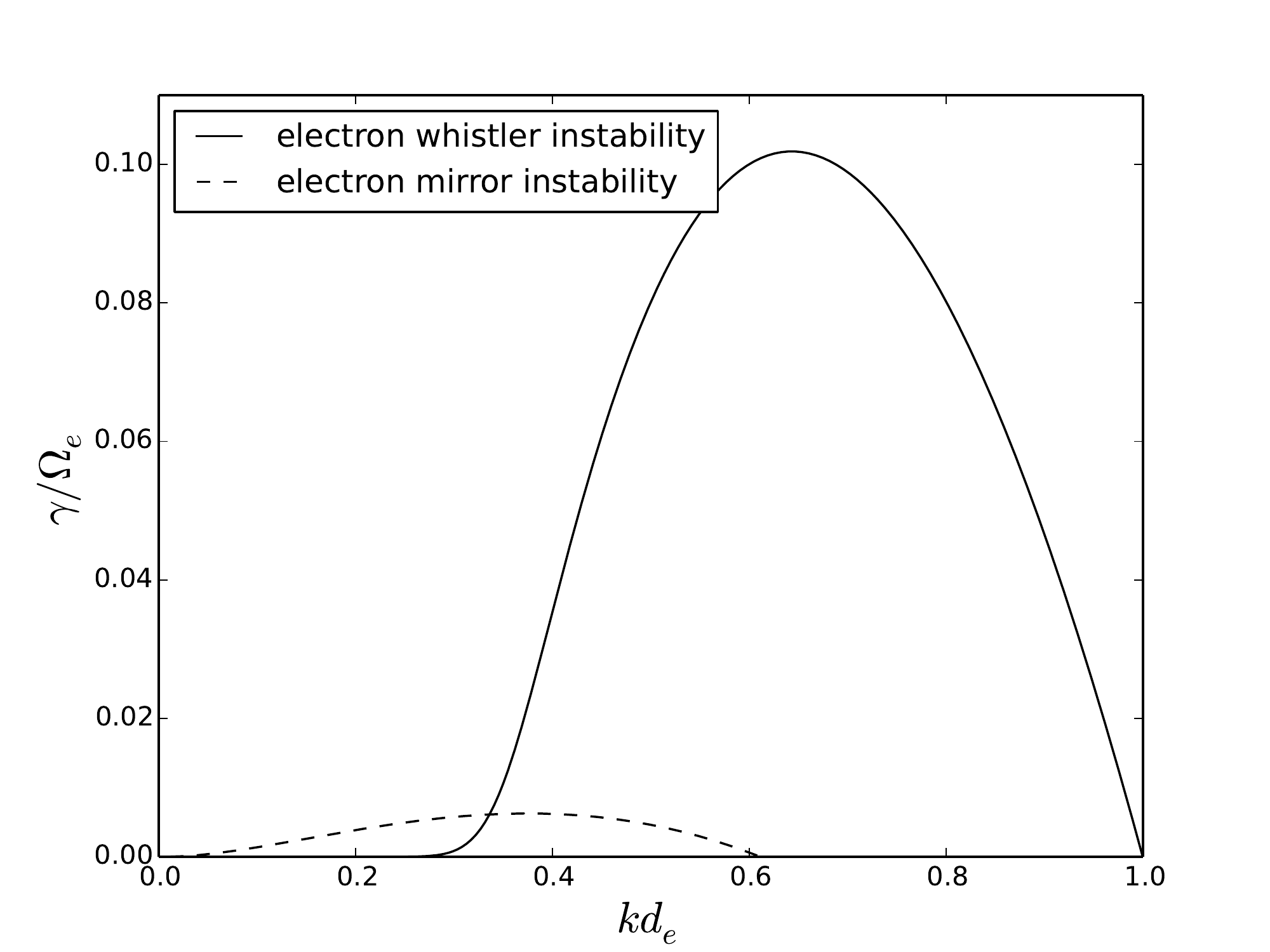}
  \caption{The growth rate as a function of wave number for
    electron whistler and electron mirror instability. Solid line shows
    the electron whistler instability growth rate at $\theta_m=0$ while
    the dashed line shows the growth rate of electron mirror
    instability at $\theta_m=73$. The maximum growth rate of electron
    whistler instability is $\gamma_m/\Omega_e = 0.1$ at $k_m d_e =
    0.64$ while the electron mirror instability maximum growth rate is
    $\gamma_m/\Omega_e= 0.006$ with $k_m d_e = 0.37$.}
  \label{fig:electrongrowthrates}
\end{figure}

\begin{figure}
  \centering
  \includegraphics[width=\columnwidth]{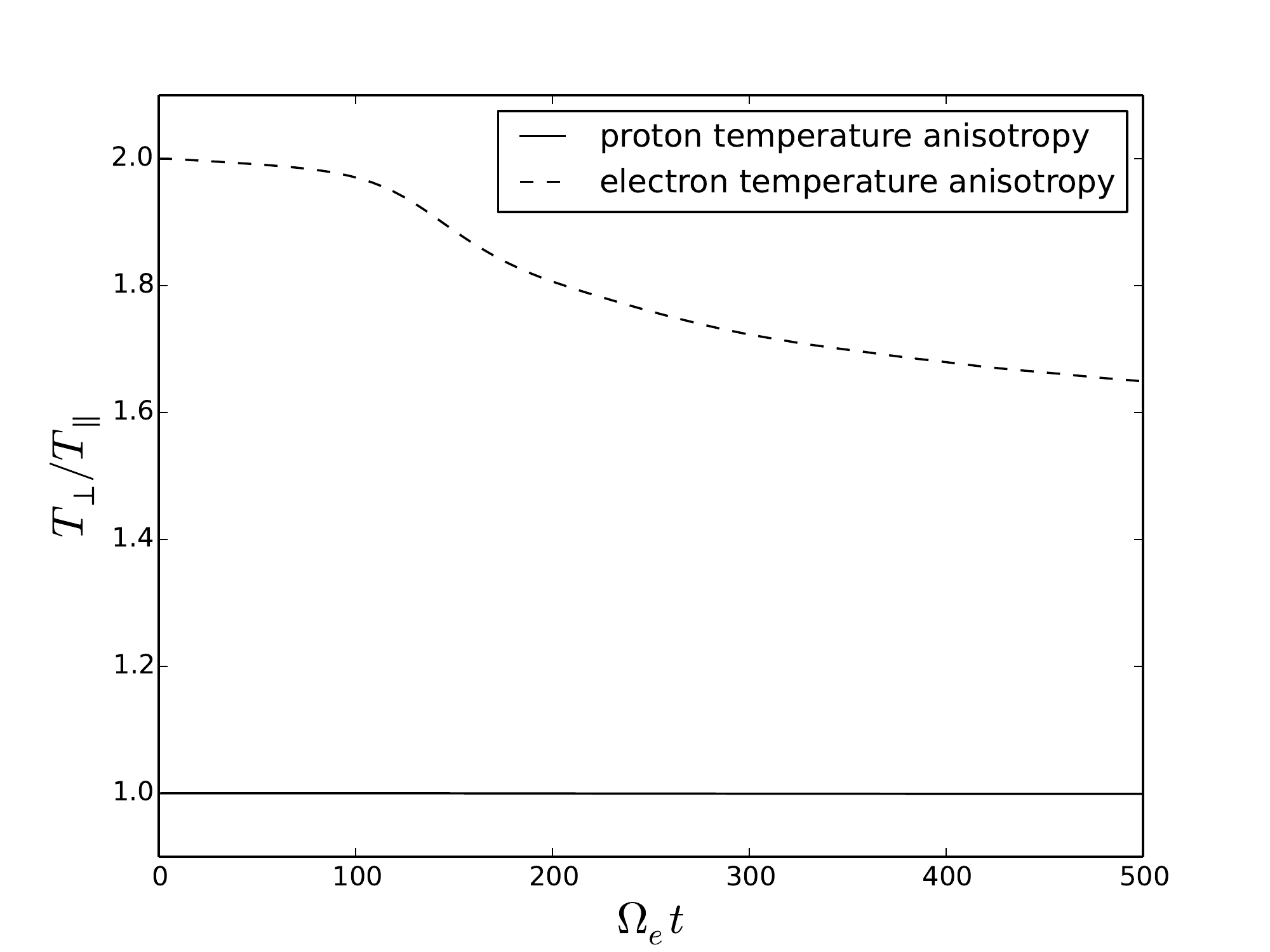}
  \caption{Proton and electron temperature anisotropy evolution
    as a function of time in 2D particle in cell simulation. Initial
    parameters are $T_{p\perp}/T_{p\parallel} = 1$,
    $T_{e\perp}/T_{e\parallel} = 2$ and $\beta_p=\beta_e=1$. The linear
    regime of proton temperature anisotropy instabilities is about
    $\Omega_e t = 100$.}
  \label{fig:electrons_aniso_benchmark}
\end{figure}

\begin{figure}
  \centering
  \includegraphics[width=\columnwidth]{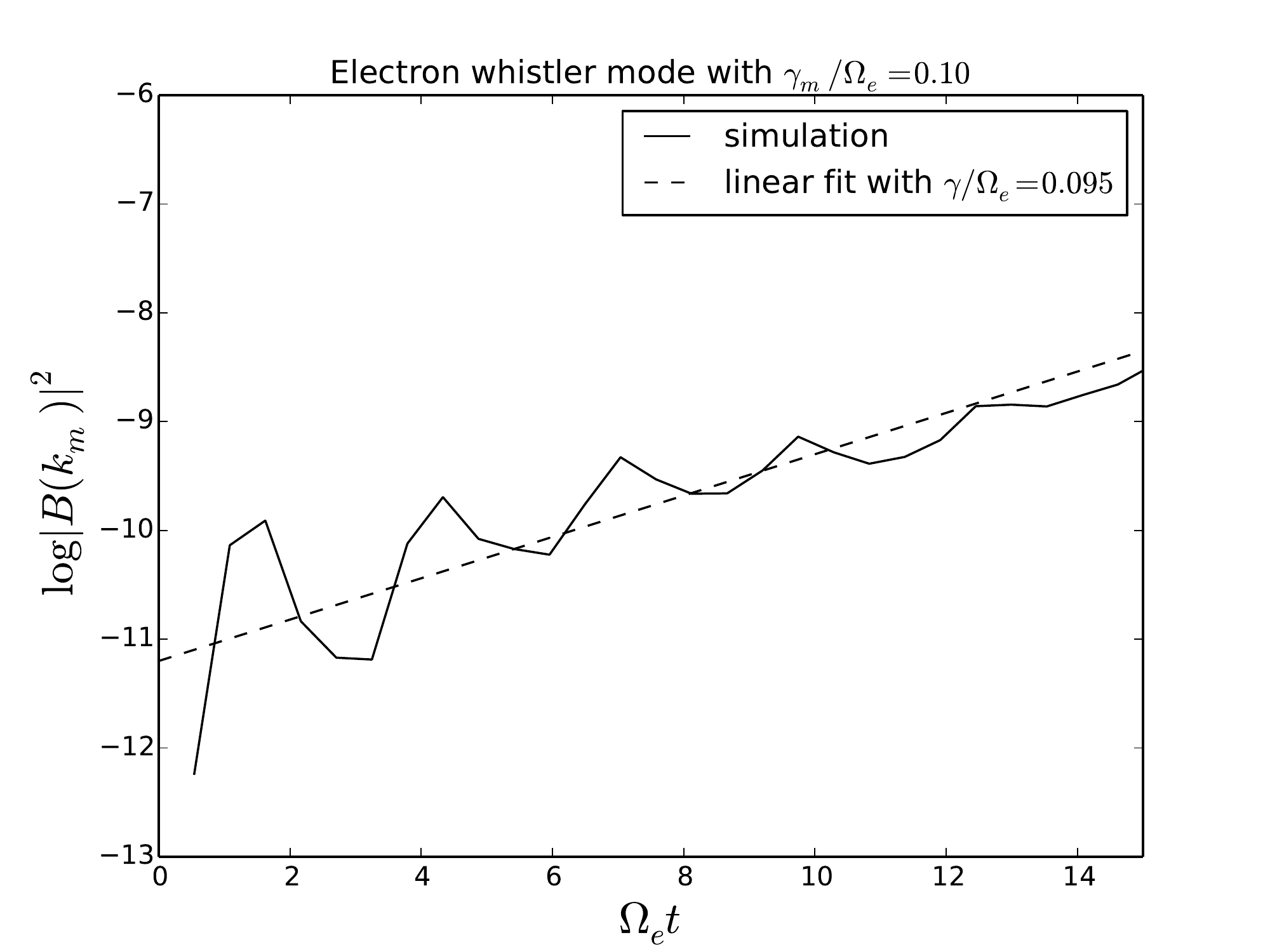}
  \caption{Measured maximum growth rate of electron whistler instability from
    simulation in the linear regime. The measured growth rate is in agreement with linear
    dispersion theory prediction. }
  \label{fig:whistlerfit}
\end{figure}

\begin{figure}
  \centering
  \includegraphics[width=\columnwidth]{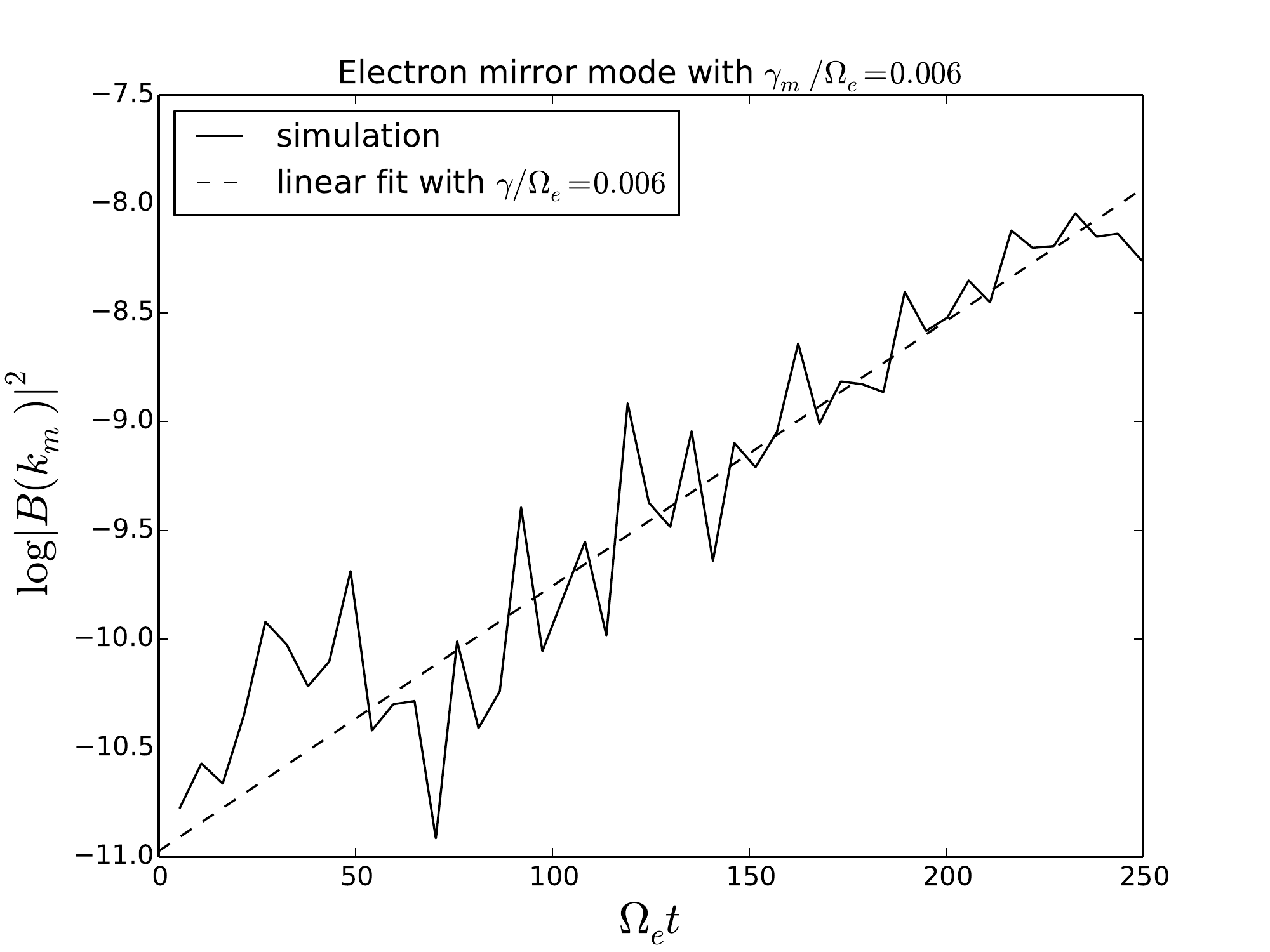}
  \caption{Measured maximum growth rate of electron mirror instability from
    simulation in the linear regime. The measured growth rate is in good agreement with linear
    dispersion theory prediction.}
  \label{fig:electronmirrorfit}
\end{figure}

We perform a similar benchmarking simulation for electron whistler and electron
mirror instabilities to show that we are resolving the electron physics in
our simulations. Here, we choose $T_{p\perp}/T_{p\parallel} =1$,
$T_{e\perp}/T_{e\parallel} =2$, $\beta_p = \beta_e = 1$, and otherwise
the same parameters as in the previous case. Now, with anisotropic
electrons $(T_{e\perp}/T_{e||}>1)$, electron whistler and
electron mirror instability grow. Figure \ref{fig:electrongrowthrates}
shows the growth rate of electron whistler and electron mirror
instabilities as a function of wave number $k$. For the given plasma
parameters, linear dispersion theory predicts that the maximum growth
rate of the electron whistler instability is $\gamma_m= 0.10\Omega_e$ at $k_m d_e =
0.64$ and $\theta=0^\circ$ while electron mirror instability has a maximum
growth rate of $\gamma_m= 0.006\Omega_e$ at $k_m d_e =
0.37$ and $\theta=73^\circ$. Figure \ref{fig:electrons_aniso_benchmark}
shows the temperature anisotropy evolution. The proton distribution stays
in equilibrium and isotropic. The electron temperature anisotropy instabilities consume the
electron free energy and isotropize the electrons. The linear regime
of the electron whistler instability lasts to about
$\Omega_et = 20$, while the linear regime of electron mirror
instability would extend to $\Omega_et = 250$ since electron mirror instability
maximum growth rate is about $17$ times smaller than the electron
whistler instability maximum growth rate. Figures
\ref{fig:whistlerfit} and \ref{fig:electronmirrorfit} show the
comparisons of the measured growth rates from simulation with linear
dispersion theory predictions. We see that the results are in a good
agreement with the predictions.

\section{Simulation Results}
We again use \texttt{PSC} to obtain the results of this section, simulating the
full nonlinear evolution of the concurrent temperature anisotropy instabilities.
First, we start with bi-Maxwellian distributions for both protons and
electrons. We choose parameters that are characteristic of the
magnetosheath.
In particular, the plasma parameters are $T_{p\perp}/T_{p\parallel} =
2.5$, $T_{e\perp}/T_{e\parallel} =1.5$, $\beta_p = 2$ and $\beta_e =
0.5$. In the magnetosheath, electrons are about $10$ times colder than
protons. We choose electron temperature to be $4$ times
colder because of the limitations of PIC simulations. We need to
resolve the electron Debye length and colder electrons means smaller
electron Debye length which needs finer grid resolutions.  We perform two-dimensional particle-in-cell simulations.
A constant background magnetic field $B_0= v_A/c=0.025$ is assumed in the
$z$ direction where $v_A$ is the proton Alfv{\'e}n speed and $c$ is speed
of light. In the magnetosheath, $v_A/c$ is about $10^{-4}$, which would lead to
unnecessarily small time steps in PIC simulations. We artificially
lower the speed of light in our simulations, to make the simulations
computationally less expensive. At the same time, we made sure to
still keep $v_A/c$ small enough to avoid introducing significant
relativistic effects.
The number of grid points $(n_y\times n_z)$
are $2048 \times 2048$. Periodic boundaries are used in each
dimension.
The number of particles used is on
average $200$ particles/cell. The size of the grid cells is taken to
be $\Delta y = \Delta z = 0.015 d_p$. 

 For these parameters, linear theory predicts, the maximum growth rate of proton
cyclotron instability to be $\gamma_m = 0.14\Omega_p$ at $k_m d_p =
0.47$, while the proton mirror instability maximum growth rate is
 $\gamma_m= 0.10\Omega_p$ with $k_m d_p = 0.53$ at 
$\theta =57^\circ$. The electron whistler instability maximum growth rate is
 $\gamma_m = 0.008\Omega_e$ with $k_m d_e = 0.6$.

Since there is an electron temperature anisotropy
($T_{e\perp}/T_{e\parallel}>1$), the electron whistler instability grows
and rapidly isotropizes the electron distribution. Also, the proton cyclotron
and the proton
mirror instability grow due to the presence of the proton temperature
anisotropy ($T_{p\perp}/T_{p\parallel}>1$).

\begin{figure}
  \centering
  \includegraphics[width=\columnwidth]{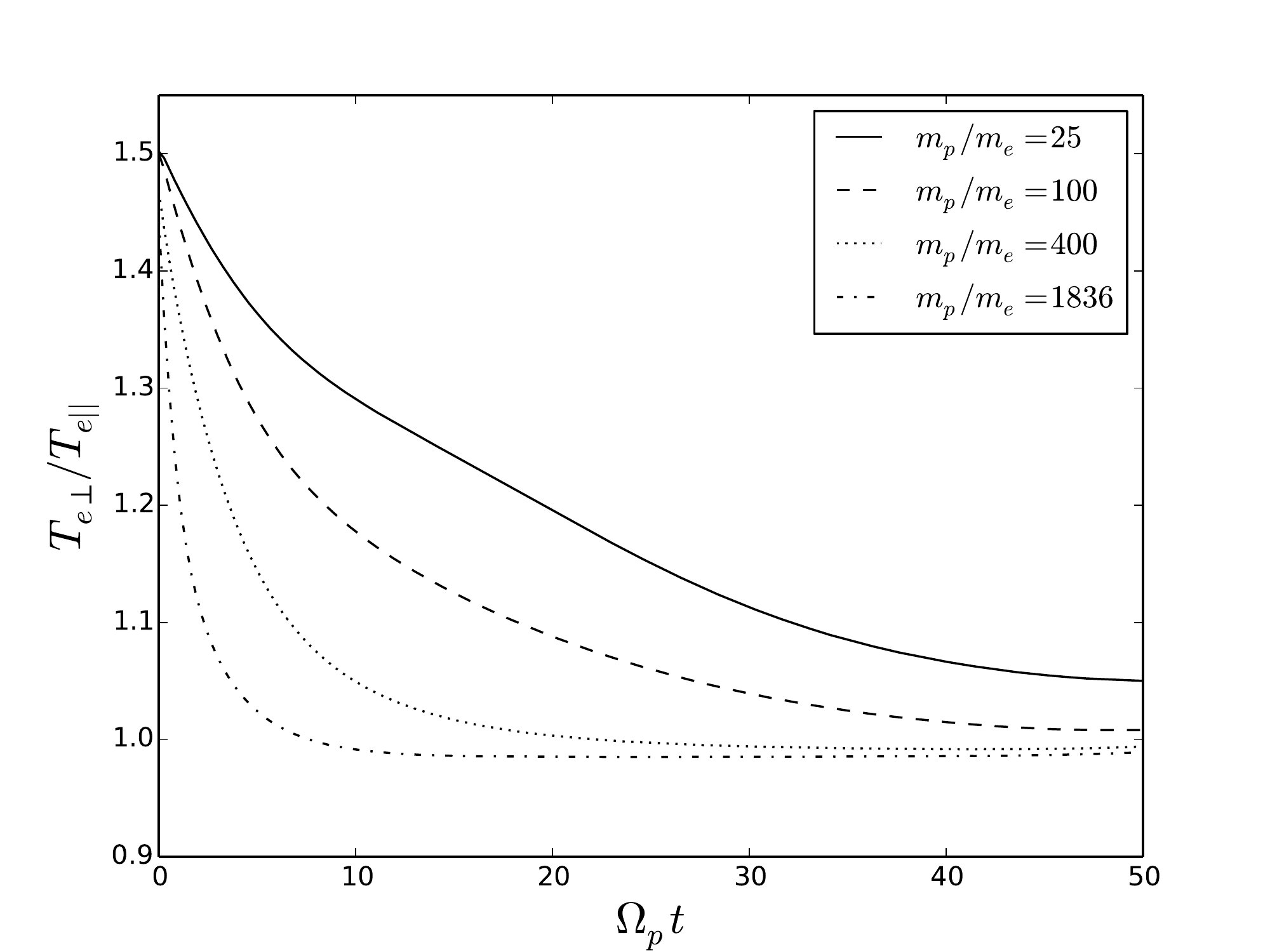}
  \caption{Electron temperature anisotropy evolution for
    different $m_p/m_e$. As we increase the mass ratio, the linear
    regime of electron whistler instability becomes smaller and
    electrons quickly isotropize.}
  \label{fig:electron_compare}
\end{figure}

\begin{figure}
  \centering
  \includegraphics[width=\columnwidth]{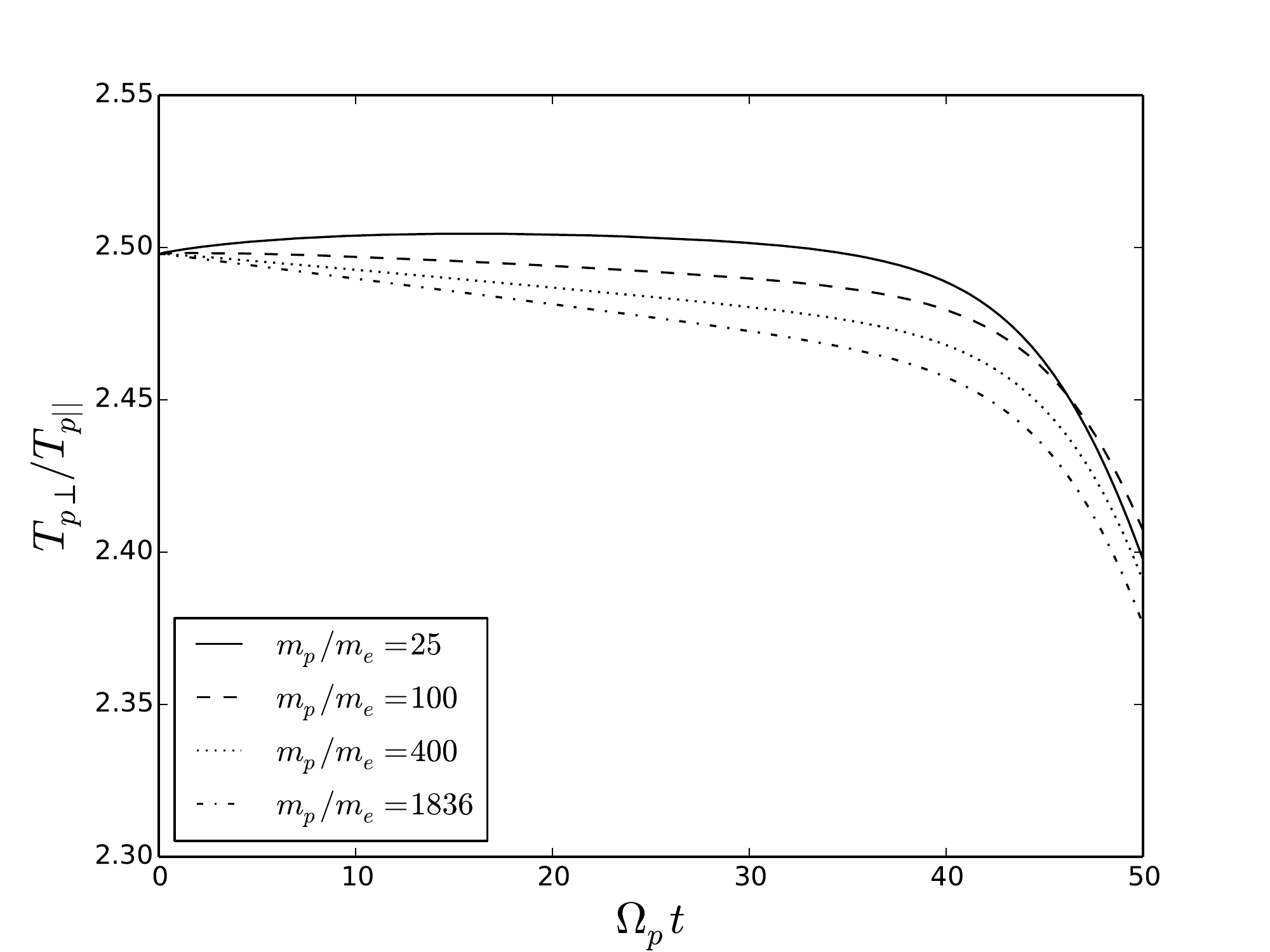}
  \caption{Proton temperature anisotropy evolution for
    different $m_p/m_e$. We only show the linear regime of proton
    instabilities. Since we are keeping $\omega_p/\Omega_p$ as a
    constant, the proton temperature anisotropy instabilties
    evolve similarly in all simulations.}
  \label{fig:proton_compare}
\end{figure}

We choose different
mass ratios $m_p/m_e = (25, 100, 400, 1836)$ and examine the electron
temperature anisotropy evolution compared to the proton temperature
anisotropy changes. Figure \ref{fig:electron_compare} shows the dependence of electron
temperature anisotropy evolution as a function of proton to electron mass ratio
($m_p/m_e$) in PIC simulations. We only show the linear regime of the proton
instabilities which lasts to about $\Omega_p t = 50$ in this case,
because we want to see how much electron temperature anisotropy is left when
the proton instabilities start to grow nonlinearly. Figure
\ref{fig:proton_compare} shows the proton temperature anisotropy as a
function of time for different $m_p/m_e$. Since we are keeping
$\omega_p/\Omega_p$ constant in all simulations, we expect the
same linear regime for proton temperature anisotropy instabilities.  Figure
\ref{fig:electron_compare} shows
that as we increase the mass ratio, the linear regime of the electron
whistler instability becomes shorter since we are making the electrons
faster and closer to reality. For $m_p/m_e = 1836$, at the end of
proton instabilities linear regime, when proton instabilities start
growing nonlinearly, there is no electron temperature anisotropy left for proton mirror
instability to take advantage of. So electron anisotropy cannot help the proton mirror
mode to dominate over the proton cyclotron instability, unless there
is a mechanism that constantly drives the electron
temperature anisotropy in the magnetosheath. The adiabatic expansion
close to the magnetopause, in the plasma depletion layer, could be a
continuous driver of the temperature anisotropies. While the
  electron distribution isotropizes more slowly at $m_p/m_e
= 25$ compared to larger mass ratios, it still leads to essentially
isotropic distributions at the end of proton instabilities linear
regime.

\begin{figure*}
  \centering
  \includegraphics[width=0.95\textwidth]{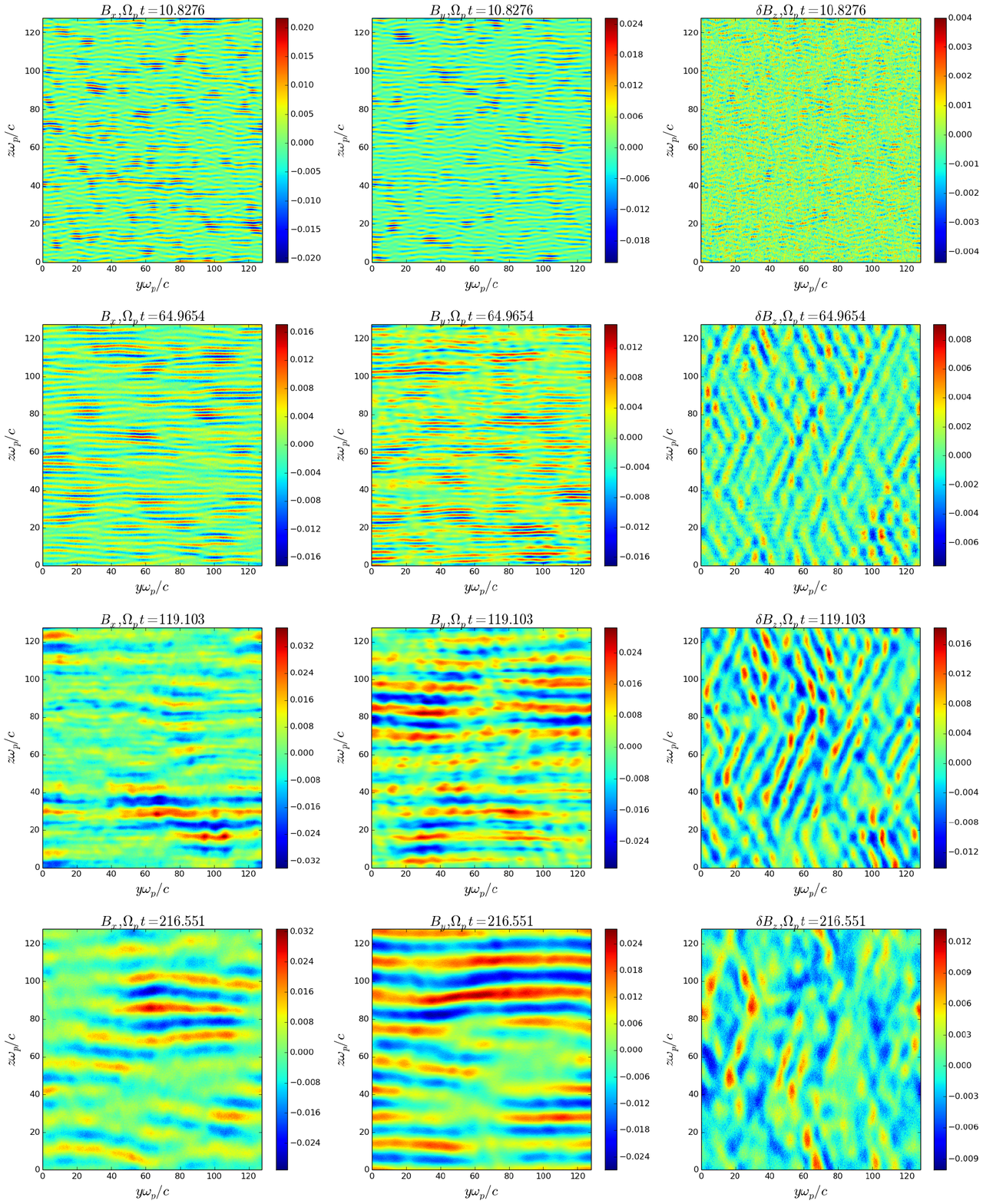}
  \caption{Time evolution of magnetic field components. First
    column shows $B_x$, second column $B_y$ and third column is $\delta
    B_z$.}
  \label{fig:B1}
\end{figure*}

In order to examine the effects of electron temperature anisotropy on
the proton mirror instability in more detail, we perform two simulations with similar
parameters and different electron temperature anisotropies. In one simulation we keep electrons isotropic and in
another one, we start with $T_{e\perp}/T_{e||} = 2$.
The simulation parameters are $T_{p\perp}/T_{p\parallel} =
2.5$, $\beta_p = 1$, $\beta_e = 1$, $m_p/m_e=25$ and $B_0 = v_A/c =
0.1$. We use $m_p/m_e = 25$ to keep the computational cost
manageable. While we have shown that by the end of the proton linear
phase, the electrons have essentially isotropized both at this as well
as at the real mass ratio, the artificially lowered mass ratio
exaggerated the effects of the electron anisotropy. This is actually
helpful as it allows us to more clearly identify the impact on the
proton instabilities.

Figure \ref{fig:B1} shows the components of the magnetic field at
different timesteps from the simulation with anisotropic electrons ($T_{e\perp}/T_{e||} = 2$). We can
see that at early timesteps, electron whistler waves gets
excited and are propagating along the background magnetic field. As
time goes on, the electron whistler instability saturates and both the
proton cyclotron and the proton mirror instability start growing. It is clear
that proton cyclotron waves are propagating along the background
magnetic field while proton mirror waves are present in the direction oblique to the background
magnetic field.

\begin{figure*}
  \centering
  \begin{minipage}{0.5\textwidth}
    \centering
    \includegraphics[width=\columnwidth]{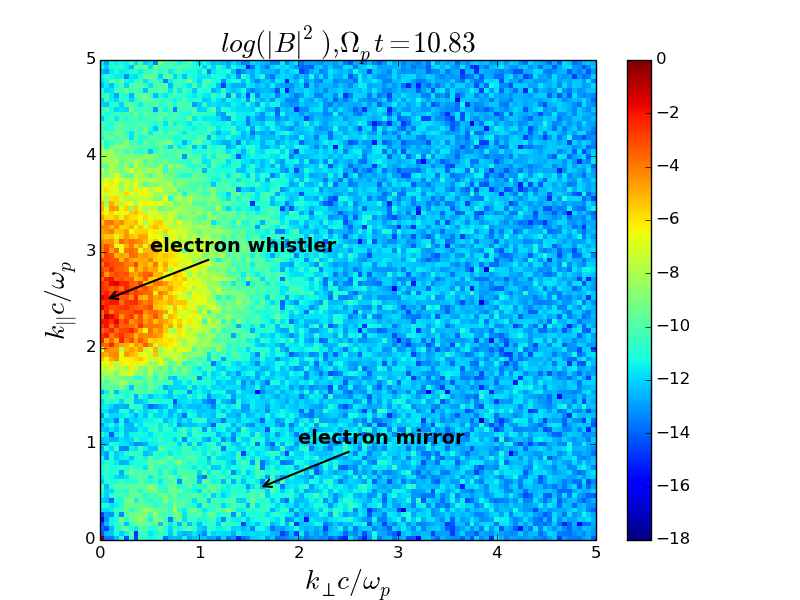}
  \end{minipage}%
  \begin{minipage}{0.5\textwidth}
    \centering
    \includegraphics[width=\columnwidth]{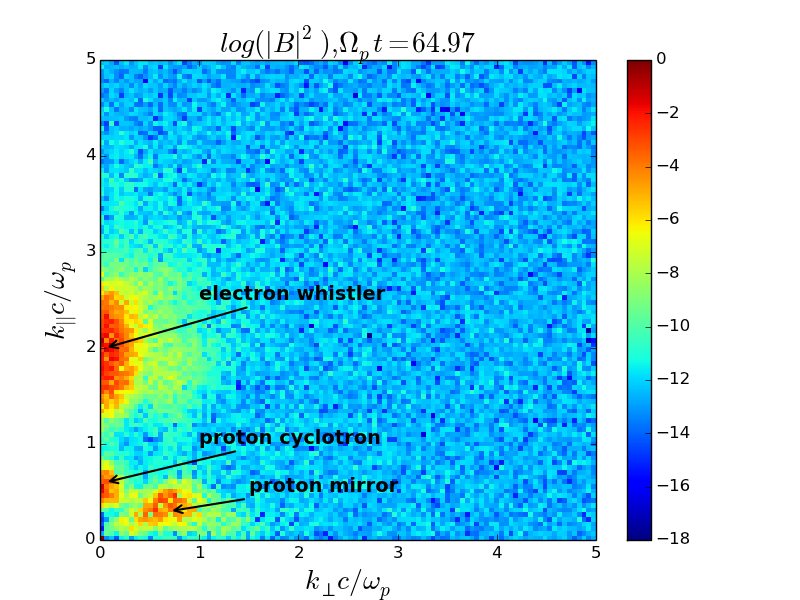}
  \end{minipage}%
  
  \begin{minipage}{0.5\textwidth}
    \centering
    \includegraphics[width=\columnwidth]{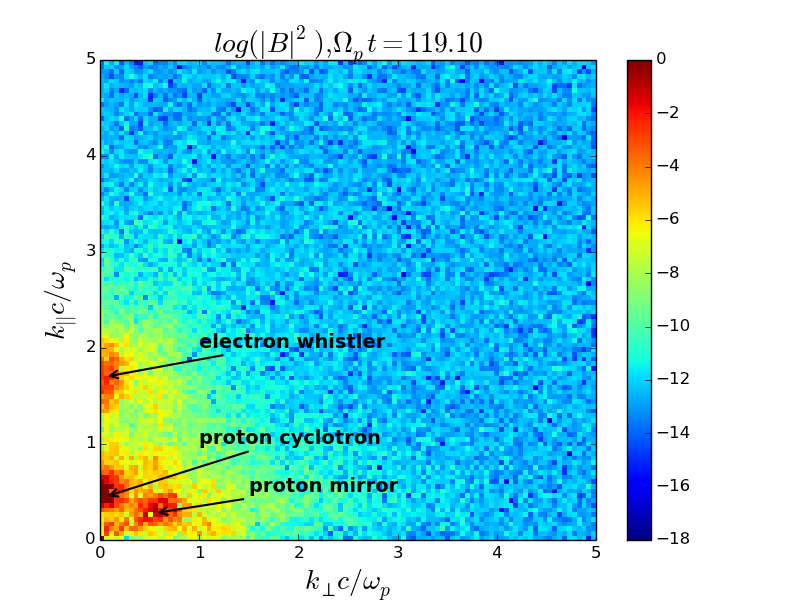}
  \end{minipage}%
  \begin{minipage}{0.5\textwidth}
    \centering
    \includegraphics[width=\columnwidth]{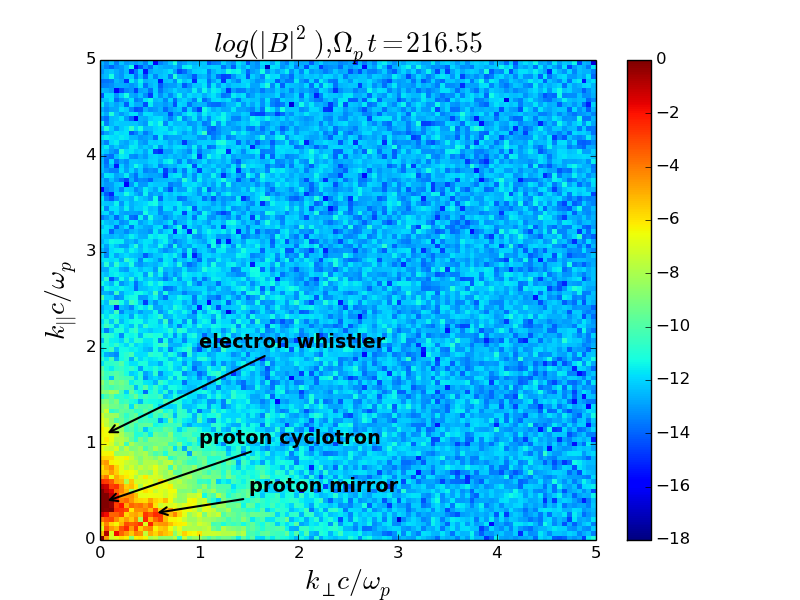}
  \end{minipage}%
   \caption{Total magnetic field spectrum. At early time,
     electron whistler instability is present. Later on, proton
     cyclotron and proton mirror instabilities grow. As each mode grows
     nonlinearly, their wavelength becomes larger and their spectrum
     moves to smaller wavenumbers.}
   \label{fig:totalBspectrum}
 \end{figure*}

Figure \ref{fig:totalBspectrum} shows the spectrum of the total magnetic
field in wavenumber space at different times. Each instability has
been marked in the
spectrum in the Figure \ref{fig:totalBspectrum}. The electron mirror
instability is about $20$ times weaker than the electron whistler
instability. At early times, the electron whistler instability is the
dominant mode. At later times, the proton cyclotron and the proton mirror
instability start growing while the electron whistler instability is still
present.

\begin{figure}
\centering
\includegraphics[width=0.5\textwidth]{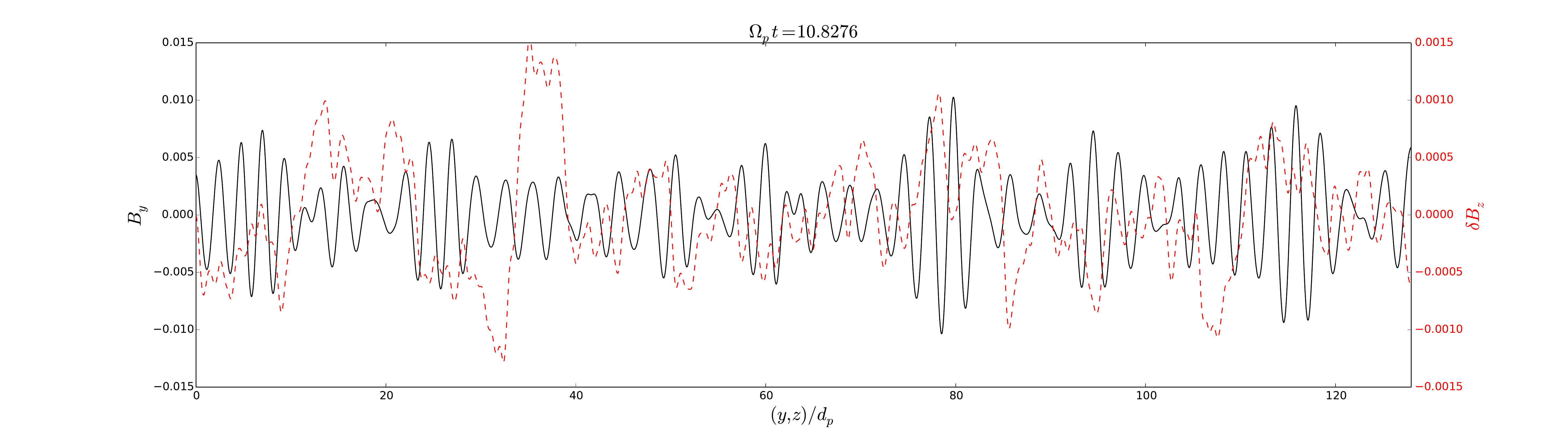}

 \includegraphics[width=0.5\textwidth]{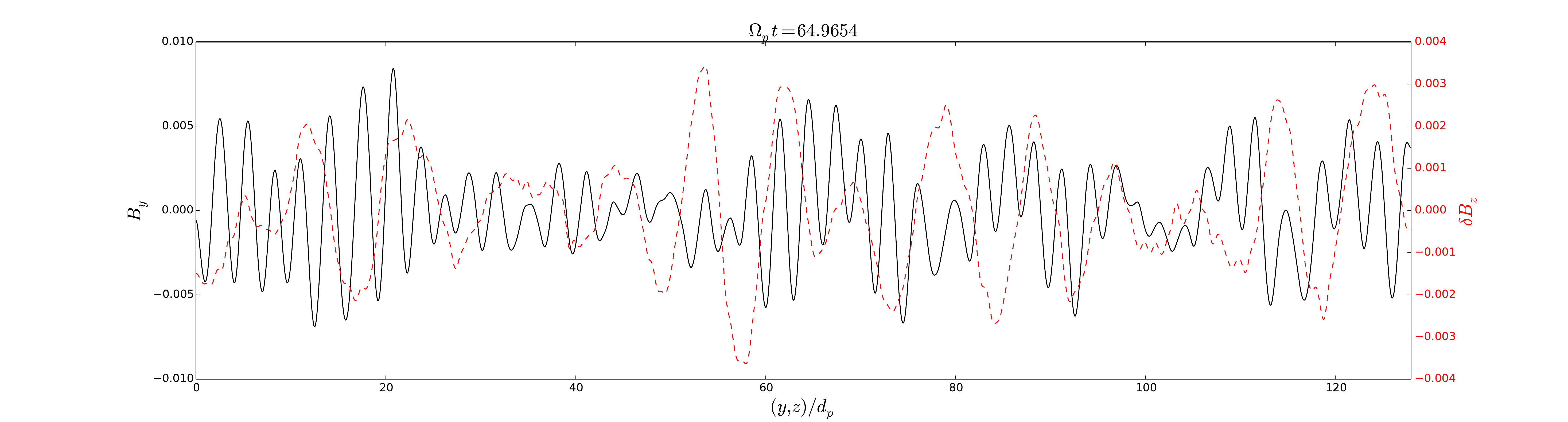}

 \includegraphics[width=0.5\textwidth]{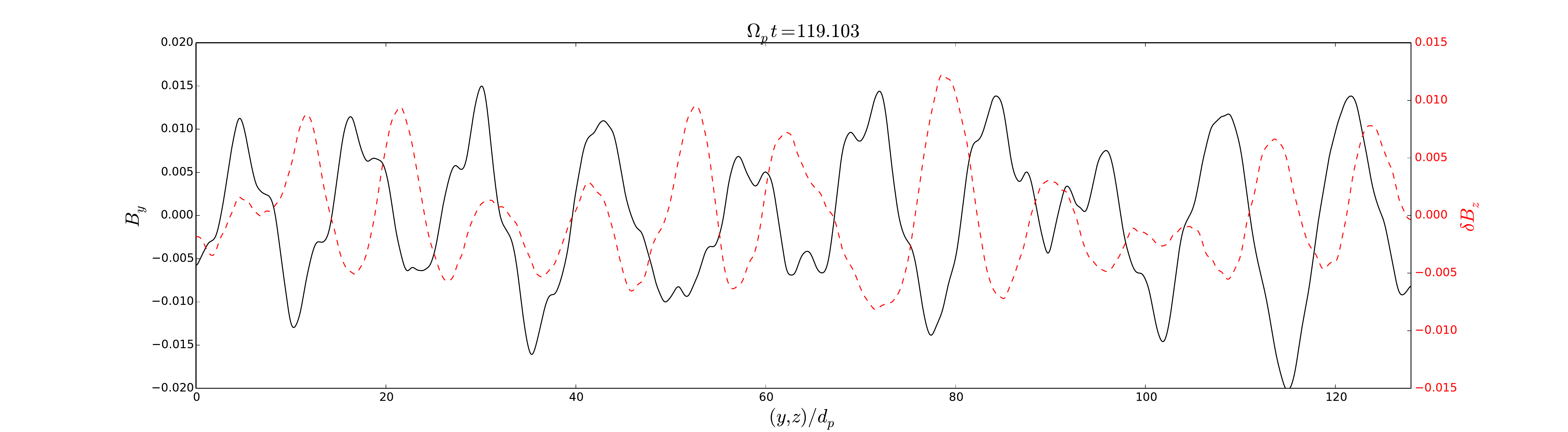}

 \includegraphics[width=0.5\textwidth]{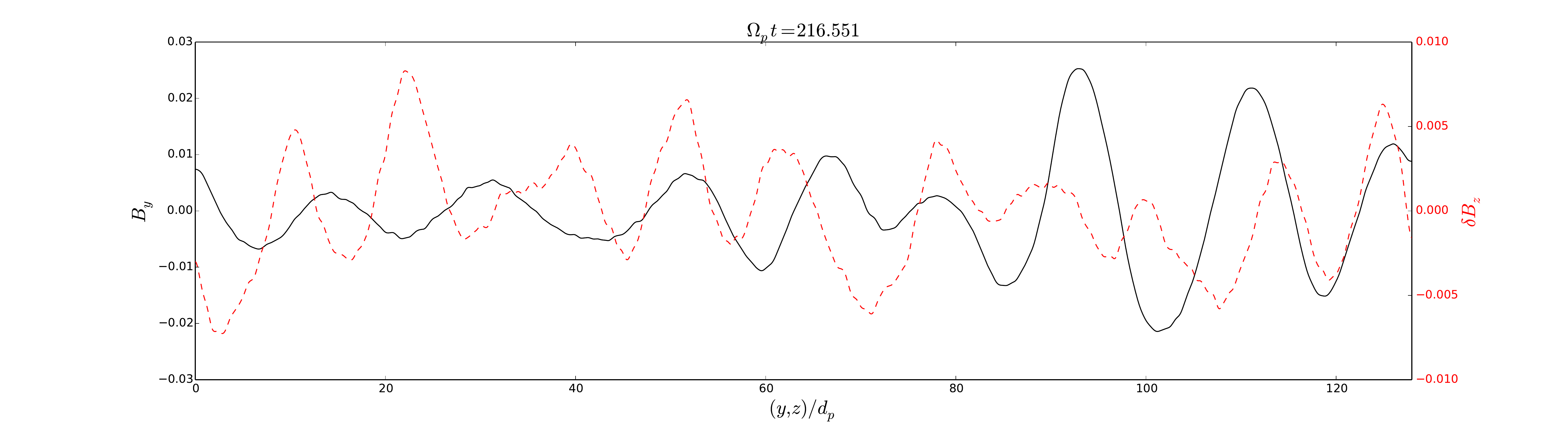}
 \caption{Magnetic field cuts of $B_y$ along $z$ direction at
   $y=64d_p$ and $\delta B_z$ along $y$
   direction at $z = 64d_p$ at different times. Black solid line shows
 $B_y$ and red dashed line is $\delta B_z$ cut.}
 \label{fig:ByBzcuts}
\end{figure}

We make cuts in the $B_y$ along $z$ direction at $y=64d_p$ and in the
$\delta B_z$ along $y$ direction at $z=64d_p$ from Figure
\ref{fig:B1}. The $B_y$ and $\delta B_z$ cuts
are shown in Figure \ref{fig:ByBzcuts}. These cuts resemble the
satellite crossings at the locations where these instabilities would
typically be present. In Figure \ref{fig:ByBzcuts}, we see the electron scale wavelengths
that are electron whistler waves and later, proton
scale wavelength structures grow which are a combination of proton
cyclotron and proton mirror mode waves. In the $\delta B_z$ cuts in Figure \ref{fig:ByBzcuts}, the proton scale structures are proton mirror
waves since proton cyclotron waves cannot have perturbations in the
direction of the background magnetic field.

\begin{figure}
  \centering
  \includegraphics[width=\columnwidth]{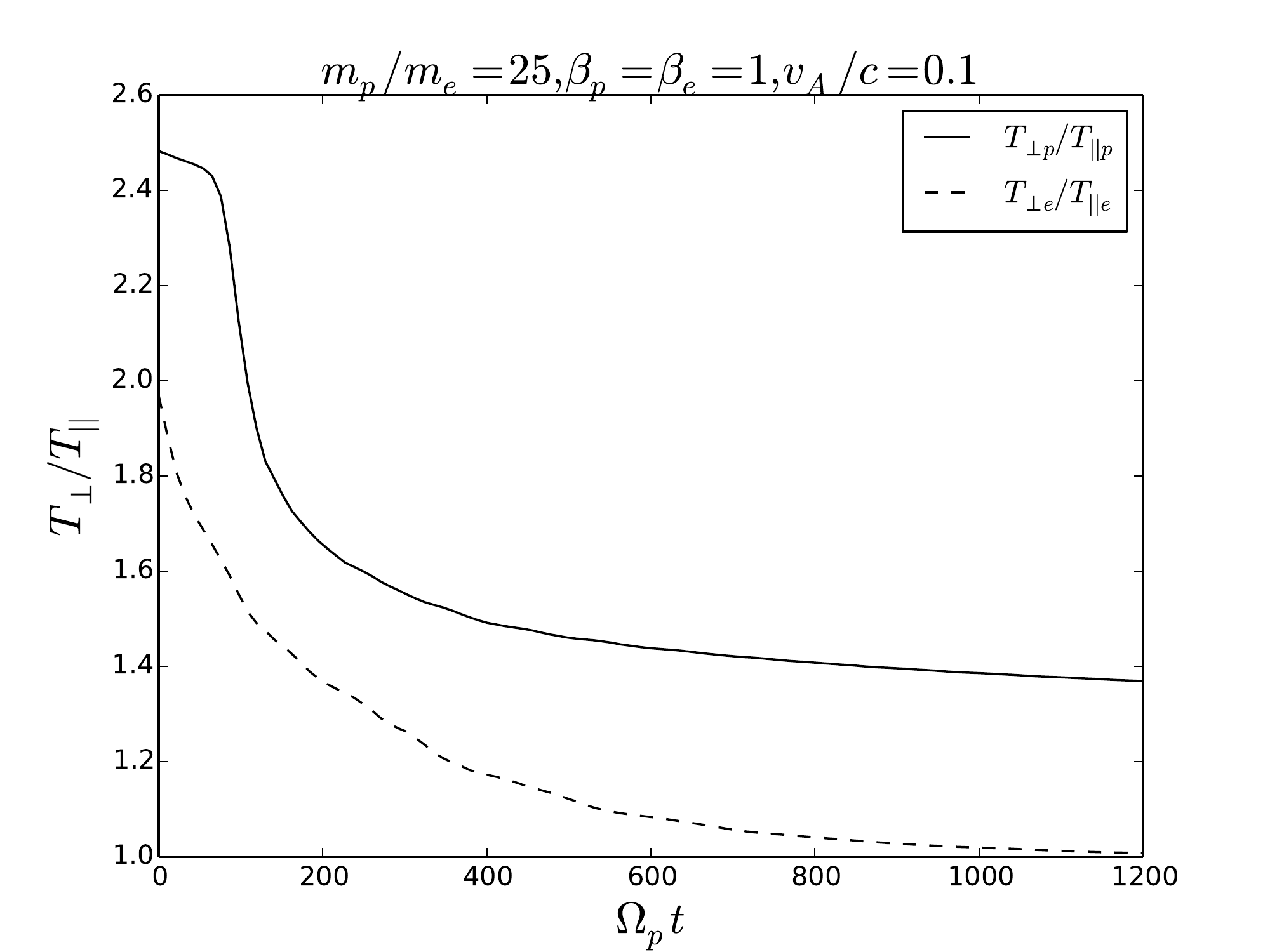}
  \caption{Temperature anisotropy evolution of protons and
    electrons with $m_p/m_e=25$.}
  \label{fig:aniso_compare}
\end{figure}

In Figure \ref{fig:aniso_compare}, the evolution of proton and
electron temperature anisotropy is shown. The proton instabilities start
growing nonlinearly around $\Omega_p t = 75$. At this time, the
electron temperature anisotropy is still $T_{e\perp}/T_{e\parallel} = 1.62$.
For plasma parameters at this timestep, the proton mirror instability is
stronger than the proton cyclotron instability. The proton cyclotron
maximum growth rate is $\gamma_m/\Omega_p = 0.07$ at $k_m d_p =
0.48$ while proton mirror instability maximum growth rate is
 $\gamma_m/\Omega_p = 0.10$ with $k_m d_p = 0.79$ at 
$\theta =62^\circ$. Then, in the nonlinear regime, both instabilities are present as shown in Figure \ref{fig:totalBspectrum}.

\begin{figure}
  \centering
  \includegraphics[width=\columnwidth]{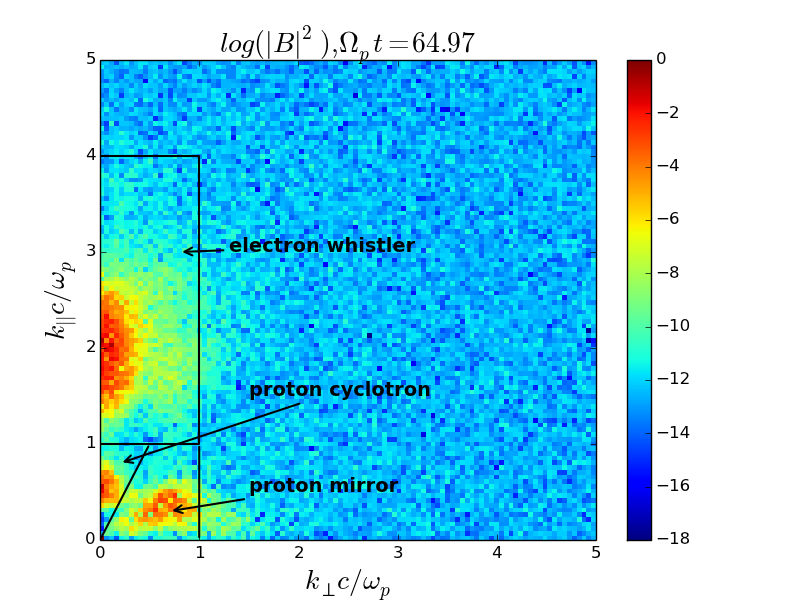}
  \caption{Energy spectrum regions for each
    instability. Electron whistler instability exist in large $k_{||}$
    region and proton cyclotron instability in small $k_{||}$. Proton
    mirror instability is present in oblique directions with $k_{\perp}
    > k_{||}$. Electron mirror instability is very weak and it doesn't
    contribute in energy density consumption.}
  \label{fig:energy_spectrum}
\end{figure}

\begin{figure}
  \centering
  \includegraphics[width=\columnwidth]{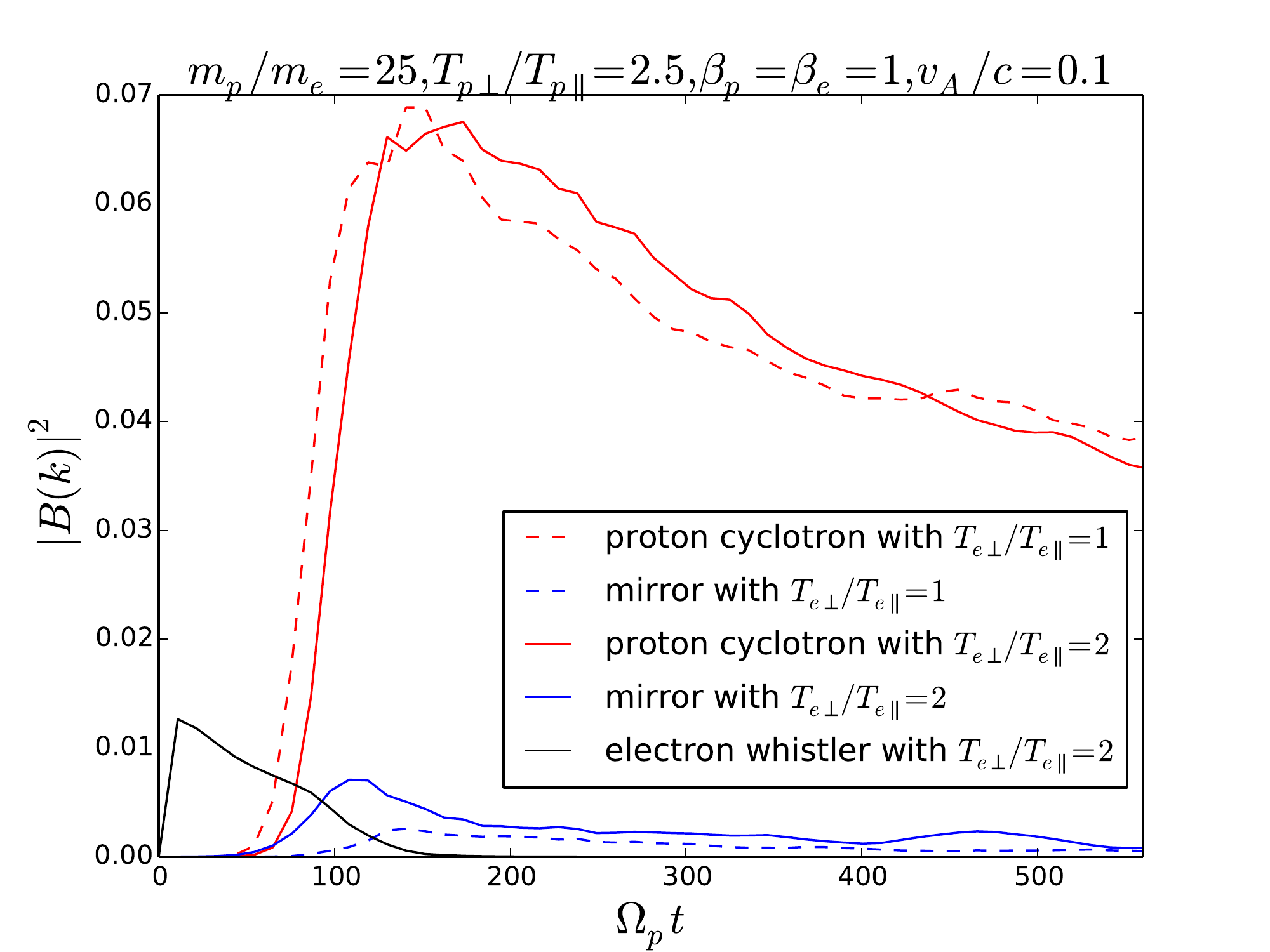}
  \caption{Energy density evolution for different
    $T_{e\perp}/T_{e\parallel}$. Solid lines show the energy density of
    the instabilities with $T_{e\perp}/T_{e\parallel}=2$ and dashed
    lines show the energy density of instabilities with
    $T_{e\perp}/T_{e\parallel}=1$. Solid black line shows the electron
    whistler instability. }
  \label{fig:energy_compare}
\end{figure}

Figure \ref{fig:energy_compare} shows the time evolution of the
magnetic energy density of proton cyclotron, proton mirror mode and electron
whistler waves. We measure the magnetic energy density of each wave by
filtering the wave spectra for each mode. The wave spectra shows three
ranges for wave number vector space as seen in Figure \ref{fig:energy_spectrum}. We define the proton cyclotron
instability range to be $0\leq \theta \leq 30^\circ$ and proton mirror
instability range is $30^\circ \leq \theta \leq 80^\circ$ for $0 <
k_{\perp,||} \leq 1$. For electron whistler instability, we choose
$0\leq \theta \leq 30^\circ$ but the wave number range is $0 <
k_{\perp} \leq 1$ and $1<k_{||} \leq 4$. 
We find a significant difference between the saturation levels of the proton
cyclotron and the proton mirror instabilities between isotropic and
anisotropic electron cases. In the isotropic case, shown in Figure
\ref{fig:energy_compare} with dashed lines, the magnetic energy
density of the proton
cyclotron instability is much larger than that of the proton mirror instability. With
isotropic electrons, the proton cyclotron instability maximum growth rate
is about 3 times stronger than the proton mirror instability, and we expect proton
cyclotron instability to consume most of the available free
energy.
In the anisotropic electrons case, the proton mirror instability maximum growth rate
is larger than that of the proton cyclotron instability, but we see that 
the magnetic energy density of the proton cyclotron
instability is still more than that of the proton mirror instability. Also,
the proton mirror instability gains more magnetic energy density
compared to the isotropic
electron case, which shows that the electron anisotropy affects the proton mirror
instability evolution. At late times, when electrons become
isotropic, the instabilities in both simulations saturate at roughly the same
magnetic energy density levels. Also, we see that the proton cyclotron
instability starts growing at a slightly different time when an electron
temperature anisotropy is present, since the presence of an electron temperature
anisotropy decreases the proton cyclotron instability growth
rate. The proton mirror instability starts growing earlier in the
anisotropic electron case, because the electron anisotropy enhances the
proton mirror instability growth rate. 

\section{Summary and Conclusion}

In this work, we have investigated the effects of electron temperature
anisotropy on the proton mirror instability evolution. Linear theory predicts
that presence of an electron temperature anisotropy can enhance the
proton mirror instability growth rate, and if it is large enough, it
can make the proton mirror instability stronger than the proton cyclotron instability. We showed that
anisotropic electrons, however, primarily drive the electron whistler
instability. 
We performed two-dimensional PIC simulations with different
electron to proton mass ratios. We studied how varying the mass ratio
affects the electron whistler instability evolution and how it impacts
the proton cyclotron and proton mirror instability growth rates.
 We find that the electron
whistler instability consumes the electron free energy before
the proton mirror instability grows into the nonlinear regime, because it grows much faster than the proton temperature anisotropy
instabilities. Therefore, all the electron free energy is gone quickly and has
little impact on the much slower proton mirror instability that has
barely started growing by that time. Our results show that
temperature anisotropy instabilities are sensitive to the chosen mass ratio
$m_p/m_e$ in PIC simulations, since an artificial mass ratio can affect the
growth and dynamics of the instabilities.

If there is a mechanism in the magnetosheath that keeps
$T_{e\perp}>T_{e||}$, it can enhance the proton mirror instability
growth rate. For example, the adiabatic expansion in the plasma
depletion layer close to the magnetopause makes
$T_{e\perp}>T_{e||}$. We will investigate this scenario further in future work.

\begin{acknowledgments}
This work was supported by National Science Foundation grant
AGS-1056898 and Department of Energy grant
DESC0006670. Computations were performed using the following resources: Trillian, a Cray
XE6m-200 supercomputer at UNH supported by the NSF MRI program under grant
PHY-1229408; XSEDE resources under contract No. TG-MCA98N022.
\end{acknowledgments}

\bibliography{article}

\end{document}